\documentclass[a4paper,11pt]{article}

\usepackage{jheppub_mod} 
\usepackage[T1]{fontenc} 

\usepackage{amsmath,booktabs}
\usepackage{graphicx}
\usepackage{xspace}
\usepackage{color}
\pagecolor{white}
\usepackage[dvipsnames]{xcolor}
\usepackage{url}
\usepackage[utf8]{inputenc}
\usepackage{epstopdf}
\usepackage{hyperref}
\usepackage{multirow}

\newcommand{\bea}{\begin{eqnarray}}
	\newcommand{\eea}{\end{eqnarray}}

\newcommand{\eq}[1]{Eq.~\eqref{#1}}

\newcommand{\beq}{\begin{equation}}
\newcommand{\eeq}{\end{equation}}

\newcommand{\GeV}{\,\text{GeV}}
\newcommand{\TeV}{\,\text{TeV}}

\newcommand{\mN}{m_N}
\newcommand{\mA}{m_A}
\newcommand{\Qwp}{Q_\text{w}^p}

\allowdisplaybreaks[1]

\preprint{CERN-TH-2021-112, PSI-21-16, ZU-TH 32/21}

\title{First-Generation New Physics in Simplified Models: From Low-Energy Parity Violation to the LHC}

\author[a,b,c]{Andreas Crivellin,}
\author[d]{Martin Hoferichter,}
\author[e,f]{Matthew Kirk,}
\author[a,b]{Claudio Andrea Manzari}
\author[g,h,i]{and Luc Schnell}

\affiliation[a]{Physik-Institut, Universit\"at Z\"urich, Winterthurerstrasse 190, 8057 Z\"urich, Switzerland}

\affiliation[b]{Paul Scherrer Institut, 5232 Villigen PSI, Switzerland}

\affiliation[c]{CERN Theory Division, 1211 Geneva 23, Switzerland}

\affiliation[d]{Albert Einstein Center for Fundamental Physics, Institute for Theoretical Physics, University of Bern, Sidlerstrasse 5, 3012 Bern, Switzerland}

\affiliation[e]{Dipartimento di Fisica, Universit\`a di Roma ``La Sapienza,'' Piazzale Aldo Moro 2, 00185 Roma, Italy}

\affiliation[f]{INFN Sezione di Roma, Piazzale Aldo Moro 2, 00185 Roma, Italy}

\affiliation[g]{LPTHE, Sorbonne Universit\'e, CNRS, 4 place Jussieu, 75252 Paris Cedex 05, France}
\affiliation[h]{Departement Physik, ETH Zürich, Otto-Stern-Weg 1, 8093 Zürich, Switzerland}
\affiliation[i]{D\'epartement de Physique, \'Ecole Polytechnique, Route de Saclay, 91128 Palaiseau Cedex, France}

\emailAdd{andreas.crivellin@cern.ch}
\emailAdd{hoferichter@itp.unibe.ch}
\emailAdd{matthew.kirk@roma1.infn.it}
\emailAdd{claudioandrea.manzari@physik.uzh.ch}
\emailAdd{luschnel@student.ethz.ch}

\abstract{New-physics (NP) constraints on first-generation quark--lepton interactions are particularly interesting given the large number of complementary processes and observables that have been measured. Recently, first hints for such NP effects have been observed as an apparent deficit in first-row CKM unitarity, known as the Cabibbo angle anomaly, and the CMS excess in $q\bar q\to e^+e^-$. Since the same NP would inevitably enter in searches for low-energy parity violation, such as atomic parity violation, parity-violating electron scattering, and coherent neutrino--nucleus scattering, as well as electroweak 
precision observables, a combined analysis is required to assess the viability of potential NP interpretations. 
In this article we investigate the interplay between LHC searches, the Cabibbo angle anomaly, electroweak precision observables, and low-energy parity violation by studying all simplified models that give rise to tree-level effects related to interactions between first-generation quarks and leptons. Matching these models onto Standard Model effective field theory, we derive master formulae in terms of the respective Wilson coefficients, perform a complete phenomenological analysis of all available constraints, point out how parity violation can in the future be used to disentangle different NP scenarios, and project the constraints achievable with forthcoming experiments.}

\begin{document}
\maketitle

\newpage
	
\section{Introduction}
\label{sec:intro}

The Standard Model (SM) of particle physics has been very successfully tested and confirmed with great precision in the last decades with the Higgs discovery in 2012 unveiling the last missing piece. While the LHC has not (yet) found any new particles directly, 
precision experiments are becoming increasingly important to gather hints as to how the SM needs to be extended, so as to eventually construct a more fundamental theory that can  account for dark matter, neutrino masses, and new sources of $CP$ violation. 
Such measurements can usually be carried out to the highest precision when particles composed of first-generation quarks and leptons are involved, given the practical life-time constraints of other SM sectors.

One class of low-energy precision experiments concerns parity violation (PV), as here QCD and QED processes that otherwise overshadow weak SM and potential NP effects are suppressed. PV is realized in atomic parity violation (APV), parity-violating electron scattering (PVES), and, more recently, coherent neutrino--nucleus scattering (CE$\nu$NS). Next, $\beta$ decays are sensitive to modifications of the charged current, with most precise constraints available from superallowed $\beta$ decays and neutron decay. Further constraints arise from electroweak precision observables (EWPO) at the $Z$ pole, and finally also at the LHC precision-frontier measurements such as non-resonant di-lepton searches are possible. This broad class of complementary measurements motivates combined analyses, especially once hints for NP arise in one or more of the processes, to elucidate which, if any, NP interpretations are viable.

Such NP hints have emerged recently from $\beta$ decays, which, in combination with kaon decays, suggest a deficit in first-row CKM unitarity, a tension referred to as the
Cabibbo angle anomaly (CAA)~\cite{Belfatto:2019swo,Grossman:2019bzp,Seng:2020wjq,Coutinho:2019aiy,Manzari:2020eum,Crivellin:2020lzu}. In addition, the CMS experiment at CERN reported a first hint for lepton flavor universality violation (LFUV) in non-resonant di-lepton searches by measuring the di-muon to di-electron ratio~\cite{Sirunyan:2021khd}. Since the CAA also permits an interpretation in terms of LFUV~\cite{Crivellin:2020lzu}, these tensions might be related to other anomalies accumulated in the flavor sector 
within the last few years. In particular, data for $b\to s\ell^+\ell^-$~\cite{Aaij:2014pli,Aaij:2014ora,Aaij:2015esa,Aaij:2015oid,Khachatryan:2015isa,ATLAS:2018gqc,CMS:2017rzx,Aaij:2017vbb}, $b\to c\tau\nu$~\cite{Lees:2012xj,Lees:2013uzd,Aaij:2015yra,Aaij:2017deq,Aaij:2017uff,Abdesselam:2019dgh}, and the anomalous magnetic moment of the muon~\cite{Bennett:2006fi,Abi:2021gix} point towards LFUV NP with a significance of $>5\sigma$~\cite{Capdevila:2017bsm,Altmannshofer:2017yso,DAmico:2017mtc,Ciuchini:2017mik,Hiller:2017bzc,Geng:2017svp,Hurth:2017hxg,Alok:2017sui,Alguero:2019ptt,Aebischer:2019mlg,Ciuchini:2019usw,Ciuchini:2020gvn}, $>3\sigma$~\cite{Amhis:2019ckw,Murgui:2019czp,Shi:2019gxi,Blanke:2019qrx,Kumbhakar:2019avh}, and $4.2\sigma$~\cite{Aoyama:2020ynm,Aoyama:2012wk,Aoyama:2019ryr,Czarnecki:2002nt,Gnendiger:2013pva,Davier:2017zfy,Keshavarzi:2018mgv,Colangelo:2018mtw,Hoferichter:2019gzf,Davier:2019can,Keshavarzi:2019abf,Kurz:2014wya,Melnikov:2003xd,Masjuan:2017tvw,Colangelo:2017fiz,Hoferichter:2018kwz,Gerardin:2019vio,Bijnens:2019ghy,Colangelo:2019uex,Blum:2019ugy,Colangelo:2014qya}, respectively.  
	
While the anomalies in semi-leptonic $B$ decays and the anomalous magnetic moment of the muon point towards NP related to second- and third-generation fermions, the
CAA and the CMS di-lepton excess can be related to first-generation quarks and leptons, with simultaneous explanations possible in terms of the effective dimension-$6$ operator $Q^{(3)}_{\ell q}$~\cite{Crivellin:2021njn}. Similarly, explanations of the CAA via modified $W$--$u$--$d$ couplings also require NP related to first generation-quarks~\cite{Belfatto:2019swo,Cheung:2020vqm,Belfatto:2021jhf,Branco:2021vhs}. In this paper, we take the large array of complementary measurements sensitive to first-generation NP, together with hints for potential NP effects, as motivation 
to perform a combined analysis, concentrating on possible correlations among the processes listed above. 

In general, however, most of these processes cannot be correlated in a model-{indepen-dent} way,\footnote{See Refs.~\cite{deBlas:2013qqa,Falkowski:2017pss} for a comprehensive analysis of 4-fermion contact interactions and Ref.~\cite{Cirigliano:2012ab} for an analysis of $\beta$ decays and LHC bounds.} due to the proliferation of independent Wilson coefficients in SM effective field theory (SMEFT)~\cite{Buchmuller:1985jz,Grzadkowski:2010es}. For this reason, we will consider a set of simplified NP models, covering the four different classes of new particles~\cite{deBlas:2017xtg} that can give rise to modified vector (or axial-vector) quark--lepton interactions below the EW symmetry breaking scale:
leptoquarks (LQs), vector bosons (VBs), vector-like quarks (VLQs),
and vector-like leptons (VLLs).
In particular, we do not consider 
particles giving rise to scalar interactions (in the $qq\ell\ell$ basis), as the resulting  currents are often related to fermion masses, and thus negligible for the first generation. Furthermore, the effect in our observables of interest would be suppressed, while other channels would be more sensitive due to chiral enhancement, e.g., $\pi\to e\nu_e/\pi\to\mu\nu_\mu$ if a charged current were generated~\cite{Bryman:2011zz,PiENu:2015seu}, and $\pi^0\to e^+e^-$ for a neutral current~\cite{Hoferichter:2021lct,KTeV:2006pwx}. The aim of our analysis is then to identify correlations among the processes described above, compare sensitivities for a given NP scenario, and project the reach of future measurements in the various search channels. 
	
In order to analyze these simplified models we will first perform the matching onto SMEFT (thus explicitly respecting $SU(2)_L$ gauge invariance) in Sec.~\ref{sec:EFT}. Then in  Sec.~\ref{sec:Obs} we express the relevant observables in terms of the Wilson coefficients of SMEFT,  including detailed discussions of the respective master formulae and NP constraints. 
The phenomenological analysis of the four classes of simplified models is performed  
in Sec.~\ref{sec:Pheno}, before we conclude in Sec.~\ref{sec:Conclusions}.
	
	\section{Standard Model Effective Theory and Simplified Models}\label{sec:EFT}
	
	As motivated in the introduction, we consider four different classes of new particles that modify quark--lepton interactions at tree level: LQs, VBs, VLQs, and VLLs. As we assume the NP to be heavy, i.e., to be realized above the EW symmetry breaking scale, we can match these models onto SMEFT. This makes $SU(2)_L$ gauge invariance explicit, and allows for a straightforward comparison of the additional relations that arise in each simplified model. In this section, 
	we will first establish our conventions for the SMEFT operator basis, before defining the NP models and performing the matching.
	
	\subsection{SMEFT}
	
	Following the conventions of Ref.~\cite{Grzadkowski:2010es}, we use the Lagrangian
	\begin{align}
		\mathcal{L} = \mathcal{L}_{\rm SM} + \sum_kC_kQ_k,
		\label{effectiveL}
	\end{align}
	where the chirality conserving dimension-6 operators (we do not consider scalar or tensor operators here) that generate 4-fermion contact interactions between quarks and leptons are
	\begin{align}
			Q_{\ell q}^{(1)ijkl} &= (\bar{\ell}_i\gamma_{\mu}\ell_j)(\bar{q}_k\gamma^{\mu}q_l), &
			Q_{\ell q}^{(3)ijkl} &= (\bar{\ell}_i\gamma_{\mu}\tau^I\ell_j)(\bar{q}_k\gamma^{\mu}\tau^Iq_l)\notag,\\
			Q_{\ell u}^{ijkl} &= (\bar{\ell}_i\gamma_{\mu}\ell_j)(\bar{u}_k\gamma^{\mu}u_l), &
			Q_{\ell d}^{ijkl} &= (\bar{\ell}_i\gamma_{\mu}\ell_j)(\bar{d}_k\gamma^{\mu}d_l),\notag\\
			Q_{q e}^{ijkl} &= (\bar{q}_i\gamma_{\mu}q_j)(\bar{e}_k\gamma^{\mu}e_l), &
			Q_{e u}^{ijkl} &= (\bar{e}_i\gamma_{\mu}e_j)(\bar{u}_k\gamma^{\mu}u_l),\notag\\
			Q_{e d}^{ijkl} &= (\bar{e}_i\gamma_{\mu}e_j)(\bar{d}_k\gamma^{\mu}d_l), & &
		\label{eq:lqOperators}
	\end{align}
	while the operators generating modified gauge-boson couplings to fermions are
	\begin{align}
			Q_{\phi q}^{(1)ij} &= (\phi^{\dagger}i\overset{\leftrightarrow}{D_{\mu}}\phi)(\bar{q}_i\gamma^{\mu}q_j), &
			Q_{\phi q}^{(3)ij} &= (\phi^{\dagger}i\overset{\leftrightarrow}{D_{\mu}^I}\phi)(\bar{q}_i\tau^I\gamma^{\mu}q_j),\notag\\
			Q_{\phi u}^{ij} &= (\phi^{\dagger}\overset{\leftrightarrow}{D_{\mu}}\phi)(\bar{u}_i\gamma^{\mu}u_j), &
			Q_{\phi d}^{ij} &= (\phi^{\dagger}\overset{\leftrightarrow}{D_{\mu}}\phi)(\bar{d}_i\gamma^{\mu}d_j),\notag\\
			Q_{\phi ud}^{ij} &= i(\tilde{\phi}^{\dagger}D_{\mu}\phi)(\bar{u}_i\gamma^{\mu}d_j). & &
		\label{eq:HiggsqOperators}
	\end{align}
	Here $q$ and $\ell$ are the quark and lepton $SU(2)_L$ doublets, while $u$, $d$, and $e$ are $SU(2)_L$ singlets. $\phi$ is the Higgs doublet, $D_\mu$ the covariant derivative, $\overset{\leftrightarrow}{D_{\mu}} = (D_{\mu} - \overset{\leftarrow}{D_{\mu}})$, $\overset{\leftrightarrow}{D_{\mu}^I} = (\tau^I D_{\mu} - \overset{\leftarrow}{D_{\mu}}\tau^I)$, and $\tau^I$ are the Pauli matrices.
	
	As we only consider first-generation fermions in this article, we can set $i,j,k,l=1$ and omit the flavor indices in the following. In general, if left-handed quarks are involved, flavor-violating $\bar u c$ and/or $\bar d s$ couplings are generated after EW symmetry breaking, which lead to various effects in flavor observables. However, by assuming alignment to the down basis, down-quark flavor-changing neutral current are avoided and only $D^0$--$\bar D^0$ mixing remains as a relevant constraint. Here the SM contributions cannot be reliably calculated, in such a way that the bounds can always be avoided if one allows for a certain degree of cancellation between SM and potential NP contributions, and we will therefore not consider flavor bounds in the following.
	
	\begin{table}\centering 
	
	\begin{tabular}{l  c c c   l  c c c} 
	\toprule
	& $SU(3)$& {$SU(2)_L$}&$U(1)_Y$ & & $SU(3)$& {$SU(2)_L$}&$U(1)_Y$\\
		\midrule
		$\Phi_1$& 3&1&$-{1}/{3}$ & $V_1$& 3&1&${2}/{3}$\\ 
		$\tilde{\Phi}_1$& 3 &1&$-{4}/{3}$ & $\tilde{V}_1$& 3 &1&${5}/{3}$\\
		$\Phi_2$&3&2&${7}/{6}$ & $V_2$&3&2&${-5}/{6}$\\
		$\tilde{\Phi}_2$&3&2&${1}/{6}$ & $\tilde{V}_2$&3&2&${1}/{6}$\\
		$\Phi_3$&3&3&$-{1}/{3}$ & $V_3$&3&3&${2}/{3}$\\
		\bottomrule
	\end{tabular}
	\caption{The ten possible scalar (vector) LQ representations $\Phi$ ($V$) under the SM gauge group. \label{LQsRep}}
\end{table}
	
	\subsection{Simplified Models and SMEFT}
	
	Extending the SM by LQs, VBs, VLQs, or VLLs gives rise to modified axial-vector or vector quark--lepton interactions described by the effective Lagrangian in \eq{effectiveL} at tree level~\cite{deBlas:2017xtg}. Here, we define LQs via their coupling to SM fermions, i.e., they have a vertex involving both a lepton and a quark. VBs are understood as QCD neutral spin-1 particles and VLQs (VLLs) are $SU(3)$ triplets (singlets) that can couple to SM quarks (leptons) via the Higgs in an $SU(2)_L$ invariant way.

	\subsubsection{Leptoquarks}
Ten LQ representations exist~\cite{Buchmuller:1986zs}, of which five are scalars ($\Phi$) and five are vectors ($V$) as given in Table~\ref{LQsRep}. Some of the LQ representations can have multiple couplings to quarks and leptons, such that overall 14 gauge-invariant interaction terms with quarks and leptons\footnote{Here we disregard couplings to two quarks, which would lead to proton decay, and can be forbidden by assigning lepton and baryon numbers to the LQs (see Refs.~\cite{Davighi:2020qqa,Greljo:2021npi} for some examples). Furthermore, we assumed that in case the LQ representation allowed for coupling to left-handed and right-handed fermions, only one of them is present at a time such that no scalar operators are generated.} are possible, as listed in Table~\ref{tab:LQ_fermion_coupling}. A complete set of LQ interactions can be found in Ref.~\cite{Crivellin:2021tmz}. 

\begin{table}
		\centering
		\begin{tabular}{ccc}
		\toprule
			{}& $\ell$&$e$\\
			\midrule
			${\bar q}$& ${\kappa_{1}^L{\gamma _\mu }V_1^{\mu} + \kappa _{3}{\gamma _\mu }\left( {\tau \cdot V_3^\mu } \right)}$&${\lambda_{2}^{LR}{\Phi _2}}$\\
			${\bar d}$& ${\tilde \lambda _{2}\tilde \Phi _2^Ti{\tau _2}}$&${\kappa_{1}^{R}{\gamma _\mu }V_1^{\mu}}$\\
			${\bar u}$& ${\lambda_{2}^{RL}\Phi _2^Ti{\tau _2}}$&${\tilde \kappa_1{\gamma _\mu }\tilde V_1^{\mu }}$\\
			${\bar q_{}^c}$& ${\lambda_{3}i{\tau _2}{{\left( {\tau \cdot{\Phi _3}} \right)}^\dag } + \lambda_{1}^{L}i{\tau _2}\Phi _1^\dag }$&$\kappa_{2}^{LR}{\gamma _\mu }{V_2^{\mu \dag }}$\\
			${\bar d_{}^c}$& ${\kappa_{2}^{RL}{\gamma _\mu }V_2^{\mu \dag} }$&${\tilde \lambda_{1}\tilde \Phi _1^\dag }$\\
			${\bar u_{}^c}$& $\tilde{\kappa}_{2}{{\gamma _\mu }\tilde V_2^{\mu \dag }}$&${\lambda_{1}^{R}\Phi _1^\dag }$\\
			\bottomrule
		\end{tabular}
		\caption{Couplings of the ten LQ representations to the SM quarks and leptons in the Lagrangian. 14 different couplings are possible, the ones of the scalar LQs are denoted by $\lambda$, the ones of the vector LQs by $\kappa$. }
		\label{tab:LQ_fermion_coupling}
	\end{table}
	The tree-level matching  onto the dimension-6 4-fermion operators in Eqs.~\eqref{eq:lqOperators} and \eqref{eq:HiggsqOperators} then yields~\cite{Alonso:2015sja,Calibbi:2015kma,Crivellin:2021egp}
	\begin{align}
			C_{\ell q}^{(1)}&= \frac{|\lambda _1^L|^2}{4m_1^2} + \frac{3|\lambda _3|^2}{4m_3^2} - \frac{|\kappa_1^L|^2}{2M_{1}^2} - \frac{3|\kappa _3|^2}{2M_3^2}, &
			C_{\ell q}^{(3)} &= -\frac{|\lambda _1^L|^2}{4m_1^2} + \frac{|\lambda _3|^2}{4m_3^2} - \frac{|\kappa_1^L|^2}{2M_{1}^2} +\frac{|\kappa _3|^2}{2M_3^2},\notag\\
			C_{\ell u} &= -\frac{|\lambda _2^{RL}|^2}{2m_2^2} + \frac{|\tilde{\kappa} _2|^2}{\tilde{M}_2^2}, &
			C_{\ell d} &= -\frac{|\tilde{\lambda} _2|^2}{2\tilde{m}_2^2} + \frac{|\kappa _2^{RL}|^2}{M_2^2},\notag\\
			C_{e u} &= \frac{|\lambda _1^R|^2}{2 m_1^2} - \frac{|\tilde{\kappa}_1|^2}{\tilde{M}_1^2}, &
			C_{e d} &= \frac{|\tilde{\lambda}_1|^2}{2\tilde{m}_1^2} - \frac{|\kappa_1^R|^2}{M_1^2},\notag\\
			C_{q e} &= -\frac{|\lambda _2^{LR}|^2}{2m_2^2} + \frac{|\kappa _2^{LR}|^2}{M_2^2}, & &
	\end{align}
where the lowercase $m$ (capital $M$) stands for the mass of the scalar (vector) LQs.
	
\subsubsection{Vector Bosons}
	
There are two possible representations under the SM gauge group for (QCD neutral) VBs that allow for couplings both to quarks and leptons: an $SU(2)_L$ singlet (triplet), denoted as $Z^\prime_\mu$ ($X^I_\mu$), both with zero hypercharge. Then the possible interactions with fermions are
	\begin{align}
	\label{Lagr_VB}
			-\mathcal{L}_{\rm VB} &= \Big(g^{\ell}\bar \ell{\gamma ^\mu }\ell+ g^e\bar e{\gamma ^\mu }e + g^d\bar d{\gamma ^\mu }d + g^u\bar u{\gamma ^\mu }u + g^q\bar q{\gamma_\mu }q\Big)Z^{\prime}_{\mu} \notag\\
			& + \left(g^{\ell}_X \bar \ell{\gamma _\mu }\frac{{{\tau^I}}}{2}\ell - g^q_X \bar q{\gamma _\mu }\frac{{{\tau^I}}}{2}q\right)X_I^\mu,
	\end{align}
	and the matching onto 2-quark--2-lepton operators in SMEFT gives
	\begin{align}
			C_{\ell q}^{(1)} &= -\frac{g^{e}g^{q}}{M_{Z^{\prime}}^2}, &
			C_{\ell q}^{(3)} &= -\frac{g^{\ell}_Xg^{q}_X}{4M_{X}^2}, &   
			C_{qe} &= -\frac{g^{q}g^{e}}{M_{Z^{\prime}}^2}, &
			C_{\ell u}^{(1)} &= -\frac{g^{\ell}g^{u}}{M_{Z^{\prime}}^2}, \notag\\
			C_{\ell d} &= -\frac{g^{\ell}g^{d}}{M_{Z^{\prime}}^2}, &
			C_{e u} &= -\frac{g^{e}g^{u}}{M_{Z^{\prime}}^2}, &
			C_{e d} &= -\frac{g^{e}g^{d}}{M_{Z^{\prime}}^2}. & 
		\label{eq:VBsingletcouplings}
	\end{align}

	\subsubsection{Vector-like Quarks}
	
	There are seven possible representations of VLQs under $SU(3)\times SU(2)_L\times U(1)_Y$, as given in Table~\ref{VLQrepresentations}.
	\begin{table}
	\centering
		\begin{tabular}{l  c c c  }
		\toprule
		& $SU(3)$& {$SU(2)_L$}&$U(1)_Y$\\
			\midrule
			$U$& 3&1&${2}/{3}$\\ 
			$D$& 3 &1&${-1}/{3}$\\
			$Q_1$&3&2&${1}/{6}$\\
			$Q_5$&3&2&${-5}/{6}$\\
			$Q_7$&3&2&${7}/{6}$\\
			$T_1$&3&3&${-1}/{3}$\\ 
			$T_2$&3&3&${2}/{3}$ \\
			\bottomrule
		\end{tabular}
	\caption{Different representations of the VLQs under the SM gauge group.}\label{VLQrepresentations}
	\end{table}
	The Lagrangian describing their interactions with SM quarks and the Higgs field is 
	\begin{align}
			-\mathcal{L}_{{\rm{VLQ}}} &= \xi^{U}\bar{U}\tilde{H}^{\dagger}q + \xi^{D}\bar{D}H^{\dagger}q +\xi^{u_1}\bar{Q}_{1}\tilde{H}u +\xi^{d_1}\bar{Q}_{1}H d \notag\\
			&+\xi^{Q_{5}}\bar{Q}_{5}\tilde{H}d+\xi^{Q_{7}}\bar{Q}_{7}H u +\frac{1}{2}\xi^{T_{1}}H^{\dagger}\tau\cdot\bar{T}_{1}q+\frac{1}{2}\xi^{T_{2}}\tilde{H}^{\dagger}\tau\cdot\bar{T}_{2}q + \text{h.c.}
	\end{align}
	Integrating out these new particles at tree level gives
	\begin{align}
	\label{eq:VLQ_SMEFT_matching}
			C_{\phi q}^{(1)} &= \frac{\left|\xi^{U}\right|^2}{4M_{U}^2} - \frac{\left|\xi^{D}\right|^2}{4M_{D}^2} - \frac{3\left|\xi^{T_1}\right|^2}{16M_{T_1}^2} + \frac{3\left|\xi^{T_2}\right|^2}{16M_{T_2}^2},  &
			C_{\phi u} &= -\frac{\left|\xi^{u_1}\right|^2}{2M_{Q_1}^2} + \frac{\left|\xi^{Q_7}\right|^2}{2M_{Q_7}^2}, \notag\\
			C_{\phi q}^{(3)} &= -\frac{\left|\xi^{U}\right|^2}{4M_{U}^2} - \frac{\left|\xi^{D}\right|^2}{4M_{D}^2} + \frac{\left|\xi^{T_1}\right|^2}{16M_{T_1}^2} + \frac{\left|\xi^{T_2}\right|^2}{16M_{T_2}^2}, &
			C_{\phi d} &= \frac{\left|\xi^{d_1}\right|^2}{2M_{Q_1}^2} - \frac{\left|\xi^{Q_5}\right|^2}{2M_{Q_5}^2}, \notag\\
			C_{\phi ud} &= \frac{\xi^{d_1} \xi^{u_1 *}}{M_{Q_1}^2} , &
	\end{align}
	where we have further assumed the limit of vanishing first-generation quark Yukawa couplings~\cite{delAguila:2000rc}. 
	
	\subsubsection{Vector-like Leptons}

	There are six representations of VLLs under the SM gauge group, as given in Table~\ref{VLLrepresentations},	which couple to SM leptons and the Higgs field via
	\begin{align}
		-\mathcal{L}_\text{VLL} &= \lambda_N\, \bar{\ell}\,\tilde{\phi}\, N + \lambda_E^i\, \bar{\ell}\,\phi\, E + \lambda_{\Delta_1}\, \bar{\Delta}_1\,\phi\, e    \notag\\
		&+\lambda_{\Delta_3}\, \bar{\Delta}_3\,\tilde{\phi}\, e + \lambda_{\Sigma_0}\, \tilde{\phi}^{\dagger}\,\bar{\Sigma}_0^I\,\tau^I\,  \ell + \lambda_{\Sigma_1}\, \phi^{\dagger}\,\bar{\Sigma}_1^I\,\tau^I\,  \ell +{\rm h.c.},
	\end{align}
	and at tree level give rise to the Wilson coefficients~\cite{delAguila:2008pw,deBlas:2017xtg,Crivellin:2020ebi,Manzari:2021prf} 
	\begin{align}
		{C_{\phi \ell}^{(1)}} &= \frac{\lambda_N\lambda_N^{\dagger}}{4M_N^2} -\frac{\lambda_E\lambda_E^{\dagger}}{4M_E^2}+\frac{3}{16}\frac{\lambda_{\Sigma_0}^{\dagger}\lambda_{\Sigma_0}}{M_{\Sigma_0}^2} -\frac{3}{16}\frac{\lambda_{\Sigma_1}^\dagger\lambda_{\Sigma_1}}{M_{\Sigma_1}^2},\notag \\
		{C_{\phi \ell}^{(3)}} &= -\frac{\lambda_N\lambda_N^{*}}{4M_N^2} -\frac{\lambda_E\lambda_E^{*}}{4M_E^2} + \frac{1}{16}\frac{\lambda_{\Sigma_0}^{\dagger}\lambda_{\Sigma_0}}{M_{\Sigma_0}^2} + \frac{1}{16}\frac{\lambda_{\Sigma_1}^{\dagger}\lambda_{\Sigma_1}}{M_{\Sigma_1}^2},\notag \\
		{C_{\phi e}}&= \frac{\lambda_{\Delta_1}^{\dagger}\lambda_{\Delta_1}}{2M_{\Delta_1}^2} - \frac{\lambda_{\Delta_3}^{\dagger}\lambda_{\Delta_3}}{2M_{\Delta_3}^2}.
		\label{VLLmatch}
	\end{align}
	
				\begin{table}
		\centering
		\begin{tabular}{l c c c  } 
		\toprule
		& $SU(3)$& {$SU(2)_L$}&$U(1)_Y$\\
		\midrule
		$N$ &1 & 1 & $0$ \\
		$E$ & 1& 1 & $-1$ \\
		$\Delta_1$ & 1 & 2 & $-1/2$\\
		$\Delta_3$ & 1 & 2 &$-3/2$ \\
		$\Sigma_0$ & 1 & 3 & $0$ \\
		$\Sigma_1$& 1 & 3 & $-1$\\
		\bottomrule
		\end{tabular}
		\caption{Different representations of the VLLs under the SM gauge group.}\label{VLLrepresentations}
	\end{table}

	\section{Observables}\label{sec:Obs}
	
	In this section we summarize the observables relevant for constraining our simplified models.

	\subsection{Parity-Violating Electron Scattering}
	
	Limits on NP couplings to electrons and first-generation quarks can be extracted from data on PVES off the proton and off nuclei. In this subsection, we review the corresponding theoretical expressions, the constraints that are currently available,  and future prospects.  
	
	\subsubsection{Low-Energy Scattering}
	
	Interference between electromagnetic and weak scattering amplitudes leads to a PV asymmetry $A_e^N$ that can be measured with a longitudinally polarized electron beam incident on an unpolarized nucleon target
	\begin{align}
	\label{AeN}
		A_e^N = \frac{\sigma^+ - \sigma^-}{\sigma^+ + \sigma^-} = \frac{t G_F}{4\pi\alpha\sqrt{2}}\frac{A_V^N(t)+A_A^N(t)}{\epsilon \big[G_E^N(t)\big]^2+\eta \big[G_M^N(t)\big]^2},
	\end{align}
	where $\sigma^\pm$ represents the cross section of the helicity-dependent elastic scattering,  $t=-Q^2$ the four-momentum transfer squared, and the kinematic quantities are defined as
	\begin{equation}
	 \eta = \frac{-t}{4\mN^2},\qquad \epsilon^{-1}=1+2(1+\eta)\tan^2\frac{\theta}{2},\qquad
	 \epsilon'=\sqrt{\eta(1+\eta)}\sqrt{1-\epsilon^2},
	\end{equation}
	with scattering angle $\theta$. The quantities $A_{V/A}^N(t)$ represent the asymmetries arising from the terms in which the vector/axial-vector part of the weak current appears on the quark side, commonly parameterized 
in the effective Lagrangian 
	\begin{align}
	\label{eq:Lee}
			\mathcal{L}_{\text{eff}}^{ee}=\frac{G_{F}}{\sqrt{2}}\sum_{q=u,d,s} \Big( C^e_{1q}\big[\bar{q}\gamma^{\mu}q\big]\big[\bar{e}\gamma_{\mu}\gamma_{5}e\big]
			+ C_{2q}^{e}\big[\bar{q}\gamma^{\mu}\gamma_{5}q\big]\big[\bar{e}\gamma_{\mu}e\big] \Big),
	\end{align}
	where $C_{1q}^e$, $C_{2q}^e$ contribute to $A_{V,A}^N(t)$, respectively. Writing
	\begin{equation}
	\label{Cee}
	 C_{1q}^e=C_{1q}^{e, \text{SM}}+C^{e, \text{NP}}_{1q},\qquad
	 C_{2q}^e=C_{2q}^{e, \text{SM}}+C^{e, \text{NP}}_{2q},
	\end{equation}
    we have the SM values
    \begin{align}
    \label{CiSM}
     C_{1u}^{e, \text{SM}}&=-0.1888, & C_{2u}^{e, \text{SM}} &= -0.0352,\notag\\
     C_{1d}^{e, \text{SM}}&=C_{1s}^{e, \text{SM}}=0.3419, & C_{2d}^{e, \text{SM}} &=C_{2s}^{e, \text{SM}}=0.0249,
    \end{align}
including radiative corrections as detailed in Refs.~\cite{Erler:2013xha,Zyla:2020zbs}. The NP contributions, expressed in terms of the SMEFT Wilson coefficients defined in Eqs.~\eqref{eq:lqOperators} and \eqref{eq:HiggsqOperators}, are given by 
	\begin{align}
			C_{1u}^{e, \rm NP} &= \frac{\sqrt{2}}{4G_F} 
			\Big( C_{\ell q}^{(3)}-C_{\ell q}^{(1)}+C_{e u}+C_{qe}-C_{\ell u}
			- |V_{ud}|^2 \Big( C_{\phi q}^{(3)} - C_{\phi q}^{(1)}\Big) + C_{\phi u} \Big),\notag \\
			C_{2u}^{e, \rm NP} &= \frac{\sqrt{2}}{4G_F} 
				\Big( C_{\ell q}^{(3)}-C_{\ell q}^{(1)}+C_{e u}-C_{qe}+C_{\ell u}
				- (1-4s_W^2) \Big[|V_{ud}|^2 \Big(C_{\phi q}^{(3)} -  C_{\phi q}^{(1)}\Big) + C_{\phi u} \Big] \Big),\notag\\
			C_{1d}^{e, \rm NP} &= \frac{\sqrt{2}}{4G_F} 
				\Big( -C_{\ell q}^{(3)}-C_{\ell q}^{(1)}+C_{e d}+C_{qe}-C_{\ell d} +C_{\phi q}^{(3)} + C_{\phi q}^{(1)} + C_{\phi d} \Big),\notag\\
			C_{2d}^{e, \rm NP} &= \frac{\sqrt{2}}{4G_F} 
				\Big( -C_{\ell q}^{(3)}-C_{\ell q}^{(1)}+C_{e d}-C_{qe}+C_{\ell d} - (1-4s_W^2) \Big[ C_{\phi q}^{(3)} - C_{\phi q}^{(1)} + C_{\phi d} \Big] \Big),
	\end{align}
	where $s_W=\sin\theta_W$ is short for the weak mixing angle. 
	Next, the nucleon matrix elements are expressed in terms of form factors according to
	\begin{align}
	 \langle N(p')|\bar q \gamma^\mu q|N(p)\rangle 
&=\bar u(p')\Big[\gamma^\mu F_1^{q,N}(t) + \frac{i\sigma^{\mu\nu}q_\nu}{2\mN}F_2^{q,N}(t)\Big]u(p),\notag\\
\langle N(p')|\bar q \gamma^\mu\gamma_5 q|N(p)\rangle
&= \bar u(p')\Big[\gamma^\mu\gamma_5 G_A^{q,N}(t)+\gamma_5\frac{q^\mu}{2\mN} G_P^{q,N}(t)\Big]u(p),
	\end{align}
	where $q=p'-p$, $t=q^2$. In particular, we will write $F_i^N(t)$ for the electromagnetic form factors, $G_A^3(t)=G_A^{u,p}(t)-G_A^{d,p}(t)$ for the triplet component of the axial-vector form factor of the proton, $G_A^{q,N}(0)\equiv g_A^{q,N}$ (with $g_A^{u,p}-g_A^{d,p}=g_A=1.27641(56)$ the axial-vector coupling of the nucleon~\cite{Markisch:2018ndu}), and define the Sachs form factors
\beq
G_E^N(t)=F_1^N(t)-\eta F_2^N(t),\qquad G_M^N(t)=F_1^N(t)+F_2^N(t).
\eeq
In these conventions, the asymmetries become
	\begin{align}
	\label{AVAA}
			A^p_V(t) &= -2(2C^e_{1u}+C^e_{1d})\bigg[\epsilon \big[G_E^p(t)\big]^2+\eta \big[G_M^p(t)\big]^2\bigg]  \notag\\
			&-2(C^e_{1u}+2C^e_{1d}) \bigg[\epsilon G_E^p(t)G_E^n(t)+\eta G_M^p(t)G_M^n(t)\bigg] \notag\\
			&-2(C^e_{1u}+C^e_{1d}+C^e_{1s}) \bigg[\epsilon G_E^p(t)G_E^{s,N}(t)+\eta G_M^p(t)G_M^{s,N}(t)\bigg]\notag\\
			&-2(C^e_{1u}+2C^e_{1d}) \bigg[\epsilon G_E^p(t)G_E^{u,d}(t)+\eta G_M^p(t)G_M^{u,d}(t)\bigg],\notag\\
			A^p_A(t) & = -\epsilon' G^p_M(t)G_A^3(t)(C^e_{2u}-C^e_{2d})-\epsilon'G^p_M(t)(g_A^{u,p}+g_A^{d,p})(C^e_{2u}+C^e_{2d}) \notag\\
			&-2\epsilon'G^p_M(t)g_A^{s,N}C^e_{2s},
	\end{align}
	where 
	\beq
	G_{E/M}^{u,d}(t)=\frac{1}{3}\Big(G_{E/M}^{d,n}(t)-G_{E/M}^{u,p}(t)\Big)-\frac{2}{3}\Big(G_{E/M}^{u,n}(t)-G_{E/M}^{d,p}(t)\Big)\,,
	\eeq
	is an isospin-breaking correction~\cite{Kubis:2006cy} (while isospin breaking in the strangeness contribution has been ignored). The weak charge of the proton is then identified as
	\beq
	\Qwp =-2(2C^e_{1u}+C^e_{1d}),
	\eeq
	but its SM prediction includes further radiative corrections not yet included in Eq.~\eqref{CiSM}, leading to
	\beq
	\label{QwSM}
	\Qwp =-2(2C^e_{1u}+C^e_{1d}+0.00005)\bigg(1-\frac{\alpha}{2\pi}\bigg)=0.0710(4).
	\eeq
	This value is slightly smaller than the naive application of Eq.~\eqref{CiSM}, $\Qwp=0.0714$, and slightly larger than the reference value quoted in Ref.~\cite{Androic:2018kni}, $\Qwp=0.0708(3)$, with a difference that traces back to a small change in $C_{1u}^{e, \text{SM}}$~\cite{Zyla:2020zbs}. The adjustments in Eq.~\eqref{QwSM} include part of the $\gamma Z$ box correction from Ref.~\cite{Blunden:2012ty}. 

	The present best measurement of $\Qwp$ comes from the $Q_\text{weak}$ experiment~\cite{Allison:2014tpu,Androic:2018kni,Carlini:2019ksi} at Jefferson Lab, which measured the asymmetry at $\langle Q^2\rangle = 0.0248 \GeV^2$ and $\langle \theta \rangle = 7.90^{\circ}$, yielding~\cite{Androic:2018kni}
	\beq
	\label{Qweak}
	A_e^p=-226.5(9.3)\times 10^{-9}. 
	\eeq
	The data were analyzed setting all Wilson coefficients except for $\Qwp$ to their SM values, which, at tree level, implies
	\begin{align}
	-2(C_{1u}^e+2C_{1d}^e)&=-2(C_{1u}^e+C_{1d}^e+C_{1s}^e)=-1,\notag\\
	-(C_{2u}^e-C_{2d}^e)&=2C_{2s}^e=1-4s_W^2,\qquad C_{2u}^e+C_{2d}^e=0,
	\end{align}
	in agreement with Ref.~\cite{Androic:2018kni} (note that our $G_A^3(0)=g_A>0$ has the opposite sign). Our formulation in terms of Wilson coefficients~\eqref{AVAA} automatically accounts for the relevant short-range radiative corrections, including what is called the ``one-quark'' axial-vector contribution in Refs.~\cite{Zhu:2000gn,Liu:2007yi}.   
	
	Updating the strangeness form factor using $\mu^s=-0.017(4)$, $\langle r^2_{M,s}\rangle=-0.015(9)\,\text{fm}^2$, $\langle r^2_{E,s}\rangle=-0.0048(6) \,\text{fm}^2$~\cite{Alexandrou:2019olr}, $\Lambda_A=1.0(2)\GeV$ (corresponding to the axial radius from Ref.~\cite{Hill:2017wgb}) for the dipole scale in $G_A^3(t)$, and estimating the axial-vector couplings as $g_A^{u,p}+g_A^{d,p}=0.40(5)$, $g_A^{s,N}=-0.05(5)$~\cite{HERMES:2006jyl,Liang:2018pis,Lin:2018obj}, we extract from Eq.~\eqref{Qweak}
	\beq
	\Qwp=0.0704(47),
	\eeq
	where we have followed the same prescription for the energy-dependent part of the $\gamma Z$ box correction~
	\cite{Hall:2015loa,Blunden:2011rd,Blunden:2012ty,Gorchtein:2011mz,Rislow:2013vta} as in Ref.~\cite{Androic:2018kni}. This value is perfectly in line both with the $Q_\text{weak}$-only result from Ref.~\cite{Androic:2018kni}, $\Qwp=0.0706(47)$, and the combination with other PVES data, $\Qwp = 0.0719(45)$. 
	In our analysis, we will retain the complete master formula~\eqref{AVAA}, as the subleading terms produce some sensitivity to combinations of Wilson coefficients other than those contained in $\Qwp$.

	The uncertainty is currently dominated by experiment, with theory uncertainties estimated in Ref.~\cite{Androic:2018kni} at the level of $4.5\times 10^{-9}$ when expressed in terms of $A_e^p$. 
	In the future, the measurement of $\Qwp$ will improve considerably by the forthcoming high-precision P2 experiment at the MESA accelerator in Mainz~\cite{Becker:2018ggl}. Conducting the experiment at lower momentum transfer ($\langle Q^2\rangle = 0.006 \text{ GeV}^2$, $\langle \theta \rangle = 35^{\circ}$) to reduce the size of the $\gamma Z$ box corrections, P2 aims to measure the proton weak charge with a relative precision of  $1.83\%$, a more than three-fold improvement over $Q_\text{weak}$. At this level of precision also theory input requires further scrutiny, see the discussion in Ref.~\cite{Becker:2018ggl}, including the role of $G_{E/M}^{u,d}(t)$ and further long-range corrections~\cite{Gorchtein:2016qtl,Erler:2019rmr}, such as PV 
	$\gamma\gamma$ boxes involving a nucleon anapole moment (called ``many-quark'' contribution in Refs.~\cite{Zhu:2000gn,Liu:2007yi}). With a dedicated backward-angle measurement planned to constrain the latter, the remaining uncertainty from nucleon form factors is projected more than a factor of two below the experimental uncertainties.  
	
	Finally, PVES scattering can also be measured off nuclei, but so far results are restricted to $^{208}$Pb~\cite{Abrahamyan:2012gp,PREX:2021umo}. Future plans include $^{48}$Ca~\cite{Kumar:2020ejz} and $^{12}$C~\cite{Becker:2018ggl}, but in both cases the major motivation concerns the presently poorly understood neutron distribution in the nucleus, in such a way that it is unlikely that meaningful constraints on NP can be extracted from PVES off nuclei alone. However, in a similar way CE$\nu$NS is also sensitive to a combination of NP couplings and nuclear structure (see Sec.~\ref{sec:CEvNS}), so that improved NP constraints are expected from a combined analysis of both classes of measurements.

	\subsubsection{Atomic Parity Violation}
	
	Apart from the very challenging measurements of PVES off nuclei at low momentum transfer, nuclear weak charges can also be accessed in APV, exploiting asymmetry amplification by stimulated emission in a highly forbidden atomic transition. Experimentally, the ratio of the PV amplitude over the Stark vector transition polarizability $\beta$ is measured (with the most precise results currently available for $^{133}$Cs~\cite{Wood:1997zq,Guena:2004sq}), which then needs to be combined with atomic-theory calculations and independent input for $\beta$. The latter can be determined semi-empirically either via a measurement together with another hyperfine amplitude~\cite{Bennett:1999pd,Dzuba:2000gf} or the scalar polarizability~\cite{Dzuba:1997df,Cho:1997zz,Toh:2019iro}, leading to the recommendation $\beta=27.064(25)_\text{exp}(21)_\text{th}a_B^3$~\cite{Zyla:2020zbs} in units of the Bohr radius $a_B$, but the uncertainty of this average does not include an error inflation to account for the $2.7\sigma$ tension between the two methods. Similarly, the coefficient of the atomic structure calculation has been under debate in the literature~\cite{Derevianko:2000dt,Johnson:2001nk,Kuchiev:2002fg,Milstein:2002ai,Porsev:2009pr,Dzuba:2012kx}, with Ref.~\cite{Zyla:2020zbs} recommending $0.8977(40)$ from Ref.~\cite{Dzuba:2012kx}, which is in $1.7\sigma$ tension with the more recent $0.8893(27)$ from Ref.~\cite{Sahoo:2021thl}. 
	Finally, despite the amplified asymmetry in the atomic system, the result is still sensitive to nuclear structure input, i.e., the neutron distribution in the nucleus. In Ref.~\cite{Cadeddu:2021dqx} the recent PREX-2 measurement~\cite{PREX:2021umo} of PVES off 
	$^{208}$Pb, in combination with a correlation to $^{133}$Cs established based on density-functional methods, was used to improve this aspect of the extraction of the weak charge of $^{133}$Cs~\cite{Cadeddu:2021dqx}
	\beq
	\label{APV_Cs}
	Q_\text{w}\big(\text{$^{133}$Cs}\big)=-72.94(43),
	\eeq
	\begin{sloppypar}
	\noindent a slight shift from $Q_\text{w}(\text{$^{133}$Cs})=-72.82(42)$~\cite{Zyla:2020zbs}. Both values lie about $1.5\sigma$ below $Q_\text{w}(\text{$^{133}$Cs})=-73.71(35)$~\cite{Sahoo:2021thl}, mainly due to the difference in the atomic-structure calculation. Both values agree with the SM prediction~\cite{Zyla:2020zbs}
	\end{sloppypar}
	\begin{align}
	\label{QwAPV}
			Q_\text{w}\big(\text{$^{133}$Cs}\big) &= -2\Big[Z\big(2C_{1u}^e+C_{1d}^e+0.00005\big)
			+ N\big(C_{1u}^e+2C_{1d}^e+0.00006\big)\Big]\bigg(1-\frac{\alpha}{2\pi}\bigg)\notag\\
			&=-73.24(5),
	\end{align}
	but the pull goes into the opposite direction. In our analysis, we will use Eq.~\eqref{APV_Cs}, bearing in mind that the uncertainties might be slightly underestimated. 
	
	The above discussion shows that the current $0.6\%$ precision of the weak charge of $^{133}$Cs is becoming limited by theory, indicating that future improvements are difficult in the Cs system. However, the PV effect is enhanced by another factor of $50$ in Ra$^+$ atoms, with a Ra-based experiment under development at  
	TRI$\mu$P~\cite{Portela:2013twa} and ISOLDE~\cite{Ra:LoI}. The projected gain in sensitivity on $s_W^2$ by a factor of $5$ would correspond to a $0.1\%$ measurement of the weak charge of Ra.
	
	\subsubsection{Parity-Violating Deep Inelastic Scattering}
	\label{sec:PVDIS}
	
	PVES can also be studied in deep inelastic reactions, with the master formula~\cite{Wang:2014guo}
\begin{equation}
A_\text{PVDIS}=\frac{3G_F Q^2}{2\pi\alpha\sqrt{2}}\frac{2C_{1u}\big[1+R_C(x)\big]-C_{1d}\big[1+R_S(x)\big]+Y_3(2C_{2u}-C_{2d})R_V(x)}{5+R_S(x)+4R_C(x)},
\end{equation}
which depends on various parton distribution functions contained in $R_C$ (charm), $R_S$ (strange), $R_V$ (valence), as well as a kinematic factor $Y_3$.
In contrast to low-energy scattering, the PVDIS process is also strongly sensitive to the $C_{2q}^e$ couplings for sufficiently large $Y_3$. The most precise measurements come from the Jefferson Lab PVDIS collaboration, who measured PVDIS off a liquid deuterium target for two kinematic settings~\cite{PVDIS:2014cmd,Wang:2014guo}
	\begin{align}
			A_\text{PVDIS}^{(1)} &= 1.156\times10^{-4}[(2C_{1u}^e-C_{1d}^e)+0.348(2C_{2u}^e-C_{2d}^e)] \notag\\
			&= -91.10(3.11)(2.97)\times10^{-6},\notag\\
			A_\text{PVDIS}^{(2)} &= 2.022\times10^{-4}[(2C_{1u}^e-C_{1d}^e)+0.594(2C_{2u}^e-C_{2d}^e)] \notag\\
			&= -160.80(6.39)(3.12)\times10^{-6},
	\end{align}
	with corresponding SM predictions $A_\text{PVDIS}^{(1)}= -87.7(7) \times 10^{-6}$ and $A_\text{PVDIS}^{(2)}= -158.9(1.0) \times 10^{-6} $. The SoLID experiment at Jefferson Lab Hall A aims to improve this measurement up to a $0.8\%$ relative error on $A_\text{PVDIS}$~\cite{Zhao:2017xej}.

	\subsection{Coherent Elastic Neutrino--Nucleus Scattering}
	\label{sec:CEvNS}
	
	Low-energy PV is also accessible in CE$\nu$NS, in which a neutrino interacts with a nucleus via a neutral current, and the elastic recoil of the nucleus is measured. This  rare process was measured for the first time by the COHERENT experiment~\cite{COHERENT:2018gft,COHERENT:2020iec}, and future measurements at a number of experiments worldwide are expected to provide complementary constraints on the Wilson coefficients probed in electron scattering. In analogy to Eqs.~\eqref{eq:Lee} and \eqref{Cee}, we write  
	the relevant effective Lagrangian as
	\begin{align}
			\mathcal{L}_{\text{eff}}^{\nu_e \nu_e}=\frac{G_{F}}{\sqrt{2}}\sum_{q=u,d,s} \Big( C^{\nu_e}_{1q}\big[\bar{q}\gamma^{\mu}q\big]\big[\bar{\nu}_e\gamma_{\mu}(1-\gamma_{5})\nu_e\big]
			+ C_{2q}^{\nu_e}\big[\bar{q}\gamma^{\mu}\gamma_{5}q\big]\big[\bar{\nu}_e\gamma_{\mu}(1-\gamma_5)\nu_e\big] \Big) ,
	\end{align}
	where
	\begin{equation}
	\label{Cnunu}
	 C_{1q}^{\nu_e}=C_{1q}^{{\nu_e}, \text{SM}}+C^{{\nu_e}, \text{NP}}_{1q},\qquad
	 C_{2q}^{\nu_e}=C_{2q}^{{\nu_e}, \text{SM}}+C^{{\nu_e}, \text{NP}}_{2q},
	\end{equation}
	and the tree-level values in the SM fulfill 
	\beq
	C_{1q}^{e, \text{SM}}\big|_\text{tree}=C_{1q}^{\nu_e, \text{SM}}\big|_\text{tree}, \qquad C_{2q}^{e, \text{SM}}\big|_\text{tree}=-(1-4s_W^2)C_{2q}^{\nu_e, \text{SM}}\big|_\text{tree}.
	\eeq
The relation of the NP contributions 
to the Wilson coefficients in Eqs.~\eqref{eq:lqOperators} and \eqref{eq:HiggsqOperators} is given by 
	\begin{align}	
		C_{1u}^{\nu_e, \rm NP} &= \frac{\sqrt{2}}{4G_F} \left( C_{\ell q}^{(3)}+C_{\ell q}^{(1)}+C_{\ell u} + |V_{ud}|^2 \Big(C_{\phi q}^{(3)} - C_{\phi q}^{(1)} \Big) - C_{\phi u} \right),\notag\\
		C_{2u}^{\nu_e, \rm NP} &= \frac{\sqrt{2}}{4G_F} \left( -C_{\ell q}^{(3)}-C_{\ell q}^{(1)}+C_{\ell u} - (1-4s_W^2) \Big[|V_{ud}|^2 \Big(C_{\phi q}^{(3)} -C_{\phi q}^{(1)}\Big) + C_{\phi u} \Big] \right),\notag \\
		C_{1d}^{\nu_e, \rm NP} &= \frac{\sqrt{2}}{4G_F} \left(-C_{\ell q}^{(3)}+C_{\ell q}^{(1)}+C_{\ell d} -C_{\phi q}^{(3)} - C_{\phi q}^{(1)} - C_{\phi d} \right) ,\notag\\
		C_{2d}^{\nu_e, \rm NP} &= \frac{\sqrt{2}}{4G_F} \left( C_{\ell q}^{(3)}-C_{\ell q}^{(1)}+C_{\ell d} - (1-4s_W^2) \Big[ C_{\phi q}^{(3)} - C_{\phi q}^{(1)} + C_{\phi d} \Big] \right). 
	\label{eq:SMEFTtoCnu}
	\end{align}
	Radiative corrections to the SM values have been studied in Refs.~\cite{Barranco:2005yy,Erler:2013xha,Tomalak:2020zfh}. Following the conventions from Ref.~\cite{Erler:2013xha}, one has
	\begin{align}
	 C^{\nu_e, \text{SM}}_{1u} &=-0.1961, 
	 & C^{\nu_\mu, \text{SM}}_{1u} &=-0.1906,
	 & C^{\nu_\tau, \text{SM}}_{1u} &=-0.1877,\notag\\
	 C^{\nu_e, \text{SM}}_{1d} &=0.3539, 
	 & C^{\nu_\mu, \text{SM}}_{1d} &=0.3511,
	 & C^{\nu_\tau, \text{SM}}_{1d} &=0.3497,\notag\\
	 C^{\nu_e, \text{SM}}_{2u} & 
	 =C^{\nu_\mu, \text{SM}}_{2u}=C^{\nu_\tau, \text{SM}}_{2u}=0.5010, & & &\notag\\
	 C^{\nu_e, \text{SM}}_{2d} & 
	 =C^{\nu_\mu, \text{SM}}_{2d}=C^{\nu_\tau, \text{SM}}_{2d}=-0.5065, & & &
	\end{align}
where we included the flavor dependence of $C_{1q}^{\nu_\ell, \text{SM}}$. Since the additional corrections from $\gamma Z$ boxes and renormalization of the axial-vector current as in Eq.~\eqref{QwAPV} are absent for CE$\nu$NS, this leads to the flavor-dependent weak charges
\begin{align}
 Q_\text{w}^{\nu_\ell}=Z Q_\text{w}^{\nu_\ell, p}
 +N Q_\text{w}^{\nu_\ell, n},
\end{align}
where $Q_\text{w}^{\nu_\ell, p}=-2(2C^{\nu_\ell}_{1u}+C^{\nu_\ell}_{1d})$, $Q_\text{w}^{\nu_\ell, n}=-2(C^{\nu_\ell}_{1u}+2C^{\nu_\ell}_{1d})$, i.e., 
\begin{align}
 Q_\text{w}^{\nu_e, p}&=0.0766, & Q_\text{w}^{\nu_\mu, p}&=0.0601, &
 Q_\text{w}^{\nu_\tau, p}&=0.0513,\notag\\
 Q_\text{w}^{\nu_\ell, n}&=-1.0233.
\end{align}
These values are consistent with $Q_\text{w}^{\nu_e, p}=0.0747(34)$, $Q_\text{w}^{\nu_\ell, n}=-1.02352(25)$, and the flavor-changing corrections 
$Q_\text{w}^{\nu_e, p}-Q_\text{w}^{\nu_\mu, p}=0.01654$, 
$Q_\text{w}^{\nu_\mu, p}-Q_\text{w}^{\nu_\tau, p}=0.00876$ from Ref.~\cite{Tomalak:2020zfh}. The main difference between Refs.~\cite{Erler:2013xha,Tomalak:2020zfh} concerns the treatment of the non-perturbative uncertainties arising in $\gamma$--$Z$ mixing diagrams involving light quark loops, but this only affects $Q_\text{w}^{\nu_\ell,p}$ and thus $Q_\text{w}^{\nu_\ell}$ at well below the percent level. In either case, due to process-dependent corrections being absorbed into the definition, there is no direct correspondence to the weak charges as defined in electron scattering, in such a way that the model-independent comparison of NP constraints has to proceed in terms of the respective Wilson coefficients. 

The cross section for CE$\nu$NS takes the form~\cite{Hoferichter:2020osn}
\begin{align}
\label{CEvNS_SM}
 \frac{d\sigma^{\nu_\ell}_A}{d T}&=\frac{G_F^2\mA}{4\pi}\bigg(1-\frac{\mA T}{2E_\nu^2}-\frac{T}{E_\nu}\bigg)\big[Q_\text{w}^{\nu_\ell}\big]^2\big|F_\text{w}(\mathbf{q}^2)\big|^2\notag\\
 &+\frac{G_F^2\mA}{4\pi}\bigg(1+\frac{\mA T}{2E_\nu^2}-\frac{T}{E_\nu}\bigg)F_A(\mathbf{q}^2),
\end{align}
where $E_\nu$/$E_\nu'$ is the energy of the incoming/outgoing neutrino, $T=E_\nu-E_\nu'$ the nuclear recoil, $\mathbf{q}$ the momentum transfer, and $\mA$ the target mass. The main sensitivity to NP effects proceeds via the weak charge, which enters as a normalization for the quark vector operators, while the correction from axial-vector operators, which contribute for non-spin-zero nuclei, is sensitive to $C_{2q}^{\nu_\ell}$. The respective nuclear form factors $F_\text{w}(\mathbf{q}^2)$ and $F_A(\mathbf{q}^2)$ are discussed in detail in Ref.~\cite{Hoferichter:2020osn}. The weak form factor is normalized as $F_\text{w}(0)=1$, but the momentum dependence is, in general, still sensitive to the $C_{1q}^{\nu_\ell}$, i.e., the short-distance physics does not simply factorize into the weak charge $Q_\text{w}^{\nu_\ell}$. The axial-vector form factor decomposes as
\beq
F_A(\mathbf{q}^2)=\frac{8\pi}{2J+1}\Big(\big(g_A^{0}\big)^2 S_{00}^\mathcal{T}(\mathbf{q}^2)+g_A^0 g_A^{1} S_{01}^\mathcal{T}(\mathbf{q}^2)+ \big(g_A^1\big)^2 S_{11}^\mathcal{T}(\mathbf{q}^2)\Big),
\eeq
where $J$ is the spin of the nucleus, the Wilson coefficients factorize into 
\beq
g_A^0=\big(C_{2u}^{\nu_\ell}+C_{2d}^{\nu_\ell}\big)\big(g_A^{u,p}+g_A^{d,p}\big)+2C_{2s}^{\nu_\ell}g_A^{s,N},\qquad 
g_A^1=\big(C_{2u}^{\nu_\ell}-C_{2d}^{\nu_\ell}\big)g_A,
\eeq
and the nuclear form factors $S_{ij}^\mathcal{T}$ are defined in such a way that the one-body contribution is normalized to
\beq
F_A(0)\big|_\text{1b}=\frac{4}{3}\frac{J+1}{J}\Big[(g_A^0+g_A^1)\langle \mathbf{S}_p\rangle+(g_A^0-g_A^1)\langle \mathbf{S}_n\rangle\Big]^2. 
\eeq
Accordingly, this response only contributes to odd-$A$ nuclei, for which the spin expectation values $\langle \mathbf{S}_N\rangle$ can be non-vanishing, but even then $F_A(\mathbf{q}^2)$
only matters for precision measurements, since the vector contribution is enhanced by a factor $\sim N^2$ due to the coherent sum over the nucleus. To reduce the nuclear uncertainties subsumed into $F_\text{w}(\mathbf{q}^2)$, either measurements need to be performed at very small momentum transfer, or improved nuclear-structure calculations are required, whose most uncertain part, again related to the neutron distribution in the nucleus, could be constrained in future global analyses of CE$\nu$NS and PVES off multiple nuclear targets. At present, ab-initio calculations are becoming available for medium-size nuclei~\cite{Payne:2019wvy}, while currently most calculations are based on (relativistic) mean field methods~\cite{Horowitz:2003cz,Patton:2012jr,Yang:2019pbx,Co:2020gwl,VanDessel:2020epd} and the large-scale nuclear shell model~\cite{Klos:2013rwa,Hoferichter:2016nvd,Hoferichter:2018acd,Hoferichter:2020osn}. 

The NP constraints that can be derived from the CE$\nu$NS measurements by COHERENT for CsI~\cite{COHERENT:2018gft} and Ar~\cite{COHERENT:2020iec} are not yet competitive, see, e.g., Refs.~\cite{Altmannshofer:2018xyo,Skiba:2020msb}, but future improvements are projected to even become interesting probes of the weak mixing angle~\cite{Canas:2018rng,AristizabalSierra:2021uob}. In our analysis, we will thus first use the present COHERENT constraints, whose neutrino beam, through 
the decay of the $\pi^+$, consist of a mixture of $\nu_\mu$, $\nu_e$, and $\bar{\nu}_\mu$. While flavor oscillations can be neglected on the scale of the experiment, this implies that only about a third of the events are actually sensitive to the first-generation NP operators, which will be taken into account in the respective exclusion plots.  
The current uncertainties on the total cross section, both for the CsI and the Ar measurement, are at the level of $30\%$, but in view of the broad CE$\nu$NS program worldwide, we also consider an optimistic projection of $1\%$-precision for a COHERENT-like decomposition of the neutrino beam. At this level of precision the subtleties regarding radiative corrections discussed above will also become relevant.

	\subsection{Beta Decays}
	\label{sec:beta_decay_CAA}
	
In addition to the neutral current probed in electron and neutrino scattering off quarks, the same effective operators also contribute to $\beta$ decays via $SU(2)_L$ invariance. In this case, observables depend on the CKM matrix element $V_{ud}$, so to be able to extract NP constraints other processes that, indirectly, enter the determination of CKM parameters become important. Here, we briefly review the relevant processes and summarize the input we will use. 

Superallowed $\beta$ decays remain the primary source of information on $V_{ud}$~\cite{Hardy:2020qwl}. While the data base, an average over a large number of nuclear decays, has been stable for many years, the accuracy of the extraction of $V_{ud}$ critically depends on hadronic~\cite{Marciano:2005ec,Seng:2018yzq,Seng:2018qru,Czarnecki:2019mwq,Seng:2020wjq,Hayen:2020cxh,Shiells:2020fqp} and nuclear~\cite{Miller:2008my,Miller:2009cg,Gorchtein:2018fxl,Hardy:2020qwl} corrections. The former can be addressed in combination with lattice-QCD calculations~\cite{Seng:2020wjq}, but improved control of the nuclear uncertainties will require a concerted effort that exploits recent advances in ab-initio nuclear-structure calculations~\cite{Gysbers:2019uyb,Martin:2021bud,Glick-Magid:2021xty}. Alternatively, $V_{ud}$ can be extracted from neutron decay~\cite{Czarnecki:2018okw,Gorchtein:2021qdf}, in which case the additional nuclear uncertainties are absent, leaving the same hadronic effects as for superallowed $\beta$ decays. In this case, significant improvements in precision measurements of the neutron life time $\tau_n$~\cite{Pattie:2017vsj,UCNt:2021pcg} and the asymmetry parameter $\lambda$~\cite{Markisch:2018ndu} have been achieved in the last years, so that, with Ref.~\cite{UCNt:2021pcg} for $\tau_n$, a gain of another factor $2$ in precision in $\lambda$ would render the neutron-decay extraction competitive with superallowed $\beta$ decays. Finally, pion $\beta$ decay $\pi^\pm \to \pi^0 e^\pm \nu_e$~\cite{Pocanic:2003pf,Cirigliano:2002ng,Czarnecki:2019iwz} is currently not measured sufficiently precisely to impose meaningful constraints, but might become relevant in the future~\cite{PIENUXe} in particular in combination with $K_{\ell 3}$ decays~\cite{Czarnecki:2019iwz}.

The confrontation of the resulting value for $V_{ud}$ with CKM unitarity is then further complicated by a tension between $K_{\ell 2}$ and $K_{\ell 3}$ decays, leading to contradicting values for $V_{us}$. For $K_{\ell 2}$ decays, traditionally, the ratio to the pion decay is studied, in such a way that only the ratio of decay constants~\cite{Dowdall:2013rya,Carrasco:2014poa,Bazavov:2017lyh} and radiative corrections~\cite{Cirigliano:2011tm,DiCarlo:2019thl} are required for the interpretation. Likewise, $K_{\ell 3}$ decays require input for the hadronic form factor $f_+(0)$~\cite{Carrasco:2016kpy,FermilabLattice:2018zqv}, but in contrast to $K_{\ell 2}$ radiative corrections~\cite{Cirigliano:2001mk,Cirigliano:2004pv,Cirigliano:2008wn,Seng:2021boy,Seng:2021wcf} have not yet been independently verified in lattice QCD. Finally, also $\tau$ decays~\cite{Amhis:2019ckw} allow for extractions of $V_{us}$, which, however, do not resolve the $K_{\ell 2}$ and $K_{\ell 3}$ tension either. 

In the end, this leads to a situation in which two tensions are present in a combined analysis of $V_{us}$ and $V_{ud}$ from kaon and $\beta$ decays, one between $K_{\ell 2}$ and $K_{\ell 3}$, the other with CKM unitarity. The exact significance depends on the choice for the various corrections described above, see Ref.~\cite{Crivellin:2020lzu} for a detailed analysis of the numerics, but a significance 
around $3\sigma$ should give a realistic estimate of the current situation. For definiteness we use the combination of Ref.~\cite{Zyla:2020zbs} for the test of CKM unitarity, where they quote:
	\begin{align}
		\big|V_{ud}\big|^2+\big|V_{us}\big|^2+\big|V_{ub}\big|^2
		&= 0.9985(5),\label{firstrow}\\
		\big|V_{ud}\big|^2+\big|V_{cd}\big|^2+\big|V_{td}\big|^2 &= 0.9970(18),
		\label{1throw}
	\end{align}
while the tensions among the kaon decays cannot be related to the first-generation NP operators we study here. In fact, even though the deficit in the first-column CKM unitarity is
	less significant than the one of the first row, it does support an interpretation in which NP affecting $V_{ud}$ could lead to an apparent violation of CKM unitarity. In view of the higher precision we will use the constraint from the first-row unitarity test in the numerical analysis.

\subsection{Electroweak Precision Observables}

EWPO include $Z$ decays measured at LEP as well as the $W$ mass (LEP+Tevatron+LHC), the Fermi constant $G_F$, the fine structure constant $\alpha$, and its running from the low to the EW scale, which depends on $\Delta\alpha_\text{had}$.\footnote{$G_F$ is taken from muon decay~\cite{MuLan:2012sih}, for $\alpha$ the tensions among determinations from $(g-2)_e$~\cite{Hanneke:2008tm,Aoyama:2019ryr} and atom interferometry~\cite{Parker:2018vye,Morel:2020dww} do not play a role, and $\Delta\alpha_\text{had}$ is taken from $e^+e^-\to\text{hadrons}$ data~\cite{Davier:2019can,Keshavarzi:2019abf}, which is robust to changes in hadronic vacuum polarization~\cite{Borsanyi:2020mff} as long as they are restricted to sufficiently low energies, see Refs.~\cite{Crivellin:2020zul,Keshavarzi:2020bfy,Malaescu:2020zuc,Colangelo:2020lcg}. } In this article we use the Python code~\texttt{smelli}~\cite{Aebischer:2018iyb,Straub:2018kue,Aebischer:2018bkb}~\texttt{v2.2.0}~\cite{smelli:v2.2.0} to perform a global fit to the EWPO (a complete list is given in Table~\ref{tab:ewpo_list}), in particular to constrain the effect of NP that modifies  $Z$ and $W$ couplings to quarks and leptons. Note that
among the set of EWPO two measurements deviate from their SM predictions by more than $2\sigma$: the $Z \to \bar b b$ forward--backward asymmetry $A_\text{FB}^{0,b}$ and the $Z \to e^+ e^-$ asymmetry parameter $A_e$. Since this later one involves first-generation fermions, it is of particular interest to our analysis, along with the ratio of the $Z$ hadronic width to $e^+ e^-$ or $\mu^+ \mu^-$ pairs ($R^0_{e,\mu}$), which also deviates from the SM, albeit with small significance of about $1.4\sigma$. Together this leads to a slight preference for non-zero NP effects for some simplified models in our phenomenological analysis.

\begin{table}[th!]
\begin{tabular}{lp{7.05cm}ll}
\toprule
Observable & Description & Exp. & Theory \\
\midrule
$\Gamma_Z$ & \small Total width of the $Z^0$ boson & \cite{Janot:2019oyi} & \cite{Freitas:2014hra,Brivio:2017vri} \\ 
$\sigma_\text{had}^0$ & \small $e^+e^-\to Z^0$ hadronic pole cross section & \cite{Janot:2019oyi} & \cite{Freitas:2014hra,Brivio:2017vri} \\ 
$R_ e^0$ & \small Ratio of $Z^0$ partial widths to hadrons vs.\ $e^+e^-$ & \cite{Janot:2019oyi} & \cite{Freitas:2014hra,Brivio:2017vri} \\ 
$R_\mu^0$ & \small Ratio of $Z^0$ partial widths to hadrons vs.\ $\mu^+\mu^-$ & \cite{Janot:2019oyi} & \cite{Freitas:2014hra,Brivio:2017vri} \\ 
$R_\tau^0$ & \small Ratio of $Z^0$ partial widths to hadrons vs.\ $\tau^+\tau^-$ & \cite{Janot:2019oyi} & \cite{Freitas:2014hra,Brivio:2017vri} \\ 
$A_\text{FB}^{0, e}$ & \small Forward--backward asymmetry in $Z^0\to  e^+ e^-$ & \cite{Janot:2019oyi} & \cite{Brivio:2017vri} \\ 
$A_\text{FB}^{0,\mu}$ & \small Forward--backward asymmetry in $Z^0\to \mu^+\mu^-$ & \cite{Janot:2019oyi} & \cite{Brivio:2017vri} \\ 
$A_\text{FB}^{0,\tau}$ & \small Forward--backward asymmetry in $Z^0\to \tau^+\tau^-$ & \cite{Janot:2019oyi} & \cite{Brivio:2017vri} \\ 
$A_ e$ & \small Asymmetry parameter in $Z^0\to  e^+ e^-$ & \cite{ALEPH:2005ab} & \cite{Brivio:2017vri} \\ 
$A_\mu$ & \small Asymmetry parameter in $Z^0\to \mu^+\mu^-$ & \cite{ALEPH:2005ab} & \cite{Brivio:2017vri} \\ 
$A_\tau$ & \small Asymmetry parameter in $Z^0\to \tau^+\tau^-$ & \cite{ALEPH:2005ab} & \cite{Brivio:2017vri} \\ 
$R_ b^0$ & \small Ratio of $Z^0$ partial widths to $ b\bar b$ vs.\ all hadrons & \cite{ALEPH:2005ab} & \cite{Freitas:2014hra,Brivio:2017vri} \\ 
$R_ c^0$ & \small Ratio of $Z^0$ partial widths to $ c\bar c$ vs.\ all hadrons & \cite{ALEPH:2005ab} & \cite{Freitas:2014hra,Brivio:2017vri} \\ 
$A_\text{FB}^{0, b}$ & \small Forward--backward asymmetry in $Z^0\to b\bar b$ & \cite{ALEPH:2005ab} & \cite{Brivio:2017vri} \\ 
$A_\text{FB}^{0, c}$ & \small Forward--backward asymmetry in $Z^0\to c\bar c$ & \cite{ALEPH:2005ab} & \cite{Brivio:2017vri} \\ 
$A_ b$ & \small Asymmetry parameter in $Z^0\to b\bar b$ & \cite{ALEPH:2005ab} & \cite{Brivio:2017vri} \\ 
$A_ c$ & \small Asymmetry parameter in $Z^0\to c\bar c$ & \cite{ALEPH:2005ab} & \cite{Brivio:2017vri} \\ 
$M_W$ & \small $W^\pm$ boson pole mass & \cite{Aaltonen:2013iut,Aaboud:2017svj} & \cite{Bjorn:2016zlr,Brivio:2017vri,Awramik:2003rn} \\ 
$\Gamma_W$ & \small Total width of the $W^\pm$ boson & \cite{Zyla:2020zbs} & \cite{Brivio:2017vri} \\ 
$\text{BR}(W\to  e\nu)$ & \small Branching ratio of $W\to  e\nu$, summed over neutrino flavors & \cite{Schael:2013ita} & \cite{Brivio:2017vri} \\ 
$\text{BR}(W\to \mu\nu)$ & \small Branching ratio of $W\to \mu\nu$, summed over neutrino flavors & \cite{Schael:2013ita} & \cite{Brivio:2017vri} \\ 
$\text{BR}(W\to \tau\nu)$ & \small Branching ratio of $W\to \tau\nu$, summed over neutrino flavors & \cite{Schael:2013ita} & \cite{Brivio:2017vri} \\ 
$\text{R}(W^+\to cX)$ & \small Ratio of partial width of $W^+\to cX$, $X=\bar d, \bar s, \bar b$ over the hadronic $W$ width & \cite{Zyla:2020zbs} & \cite{Brivio:2017vri} \\ 
$\text{R}_{\mu  e}(W\to \ell\nu)$ & \small Ratio of branching ratio of $W\to \mu\nu$ and  $W\to  e\nu$, individually summed over neutrino flavors & \cite{Schael:2013ita,Aaij:2016qqz} & \cite{Brivio:2017vri} \\ 
$\text{R}_{\tau  e}(W\to \ell\nu)$ & \small Ratio of branching ratio of $W\to \tau\nu$ and  $W\to  e\nu$, individually summed over neutrino flavors & \cite{Abbott:1999pk,Schael:2013ita} & \cite{Brivio:2017vri} \\ 
$\text{R}_{\tau \mu}(W\to \ell\nu)$ & \small Ratio of branching ratio of $W\to \tau\nu$ and  $W\to \mu\nu$, individually summed over neutrino flavors & \cite{Schael:2013ita,Aad:2020ayz} & \cite{Brivio:2017vri} \\ 
$A_ s$ & \small Asymmetry parameter in $Z^0\to s\bar s$ & \cite{Abe:2000uc} & \cite{Brivio:2017vri} \\ 
$R_{uc}^0$ & \small Average ratio of $Z^0$ partial widths to $u\bar u$ or $c\bar c$ vs.\ all hadrons & \cite{Zyla:2020zbs} & \cite{Freitas:2014hra,Brivio:2017vri} \\ 
\bottomrule
\end{tabular}
\caption{List of EWPO used in our analysis, along with the relevant experimental and theoretical references.}
\label{tab:ewpo_list}
\end{table}

\subsection{LHC Bounds}

\begin{table}
	\centering
	\begin{tabular}{l c c c c  } 
	\toprule
			$\mathcal{M}^\text{NP}_{q \bar{q}, AB}$ & $LL$ & $LR$ & $RL$ & $RR$\\
			\midrule
			$q = u$  & $C_{\ell q}^{(1)} - C_{\ell q}^{(3)}$  & $C_{qe}$ & $C_{\ell u}$ & $C_{eu}$ \\
			$q = d$  & $C_{\ell q}^{(1)} + C_{\ell q}^{(3)}$  & $C_{qe}$ & $C_{\ell d}$ & $C_{ed}$ \\
			\bottomrule
		\end{tabular}
		\caption{NP amplitudes $\mathcal{M}_{q \bar{q}, AB}^\text{NP}$ from the 2-quark--2-lepton operators in SMEFT.}
		\label{tab:DY2q2l}
 \end{table}

	Both the ATLAS and CMS collaboration recently presented a new non-resonant di-lepton analysis in Ref.~\cite{Aad:2020otl} and Ref.~\cite{Sirunyan:2021khd}, respectively. These data can be used to constrain the LQ and VB models as they give effects in non-resonant di-lepton events (assuming that the $Z^\prime$ is heavier than $\approx 5.5\,$TeV). To this end, we first compute the partonic cross sections 
	\begin{equation}
		\hat{\sigma}^{\text{SM+NP}}_{q \bar{q}} \equiv \hat{\sigma}^{\text{SM+NP}}(q \bar{q} \to e^+ e^-),
	\end{equation}
	with $q \in \{ u,d,s,c,b \}$ at tree level. The SM tree-level amplitudes are
	\begin{align}
			\mathcal{A}_{u_j\bar{u}_j, AB}^{\text{SM}} &= -\frac{2}{3} \frac{e^2}{q^2} - \frac{g^2}{c_W^2 q^2} \left(I_{3, A} - \frac{2}{3} s_W^2 \right) \left(I_{3, B} - s_W^2 \right),\notag\\
			\mathcal{A}_{d_j \bar{d}_j, AB}^{\text{SM}} &= \frac{1}{3} \frac{e^2}{q^2} - \frac{ g^2}{c_W^2 q^2} \left(-I_{3, A} + \frac{1}{3} s_W^2 \right) \left(I_{3, B} - s_W^2 \right),
		\label{eq:DYSM}
	\end{align}
	with $u_j \in \left \{ u,c \right \}$, $d_j \in \left \{ d,s,b \right \}$, $A,B \in \left \{L,R \right \}$, $I_{3, L} = \frac{1}{2}$, $I_{3, R} = 0$, and $e=g s_W$.
	The NP amplitudes $\mathcal{M}_{q \bar{q}, AB}^\text{NP}$ from the 2-quark--2-lepton operators are given in Table~\ref{tab:DY2q2l}.

	The partonic cross section is then proportional to
	\begin{equation}
		\hat{\sigma}^{\text{SM+NP}}_{q \bar{q}} \sim \left | M_{q \bar{q}, AB}^{\text{SM}} + M_{q \bar{q}, AB}^{\text{NP}}  \right |^2 ,
	\end{equation}
	and the total cross section is obtained by integrating over the luminosities
	\begin{equation}
		\sigma^{\text{SM+NP}} = \sum_{i} \int_{\hat{s}_\text{min}}^{\hat{s}_\text{max}} \frac{d \hat{s}}{\hat{s}} \left(\frac{d\mathcal{L}_{q\bar{q}}}{d\hat{s}}\right) \left(\hat{s} \hat{\sigma}^{\text{SM+NP}}_{q\bar{q}} \right),
		\label{eq:totalXsection}
	\end{equation}
	with the parton center-of-mass energy $\hat{s}$ equal to the square of the invariant mass of the electron--positron pairs $m^2_{e\bar{e}}$ and the differential parton--anti-parton luminosities \cite{Campbell:2006wx} given by
	\begin{equation}
			\frac{d \mathcal{L}_{q \bar{q}}}{d \hat{s}} = \frac{1}{s} \int_\tau^1 \frac{dx}{x} f_q (x, \sqrt{\hat{s}}) f_{\bar{q}} \left( \frac{\tau}{x}, \sqrt{\hat{s}}\right) + \left(q \leftrightarrow \bar{q} \right),
		\label{eq:luminosity}
	\end{equation}
	where $s$ is the beam center-of-mass energy and $\tau = \hat{s}/s$. In our numerical analysis, we use the PDF set NNPDF23LO, also employed, e.g., in the ATLAS analysis, to generate the signal Drell--Yan process~\cite{Aad:2020otl}, with the help of the \texttt{Mathematica} package \texttt{ManeParse}~\cite{Clark:2016jgm}.
	
	The CMS collaboration also provided results for the differential cross-section ratio
	\begin{equation}
		R_{\mu\bar{\mu} / e\bar{e}} \equiv \frac{d\sigma(pp \to \mu^+\mu^-)/dm_{\mu\bar{\mu}}}{d\sigma(pp \to e^+e^-)/dm_{e\bar{e}}},
	\end{equation}
	for the nine $m_{\ell \bar{\ell}}$  ($\ell = e, \mu$) bins between 200, 300, 400, 500, 700, 900, 1250, 1600, 2000, and 3500  GeV. By distinguishing the cases where zero, or at least one, of the leptons were detected in the CMS endcaps, eighteen values were obtained, denoted by $R_{\mu\bar{\mu}/e\bar{e}, i}^{\text{Data}}$ with $i = 1,\dots,18$. These were then normalized to the SM predictions obtained from Monte-Carlo simulations
	\begin{equation}
		R_{\mu\bar{\mu}/e\bar{e}, i}^{\text{Data}} / R_{\mu\bar{\mu}/e\bar{e}, i}^{\text{MC}}.
	\end{equation}

	In the double ratio, many of the experimental uncertainties cancel~\cite{Greljo:2017vvb} and, importantly, the double ratio was further normalized to unity in the first bin from 200 to 400 GeV to correct for the different detector sensitivity to electrons and muons. In order to relate the NP Wilson coefficients to the CMS measurements, we compute $R_{\mu\bar{\mu}/e\bar{e}, i}^{\text{SM+NP}}$. We then determine the best fit to data via a $\chi^2$ statistical analysis, defining
		\begin{align}
			\chi^2\equiv \sum_{i = 1, \dots,18}\dfrac{\left(\dfrac{R_{\mu\bar{\mu}/e\bar{e}, i}^{\text{Data}}}{R_{\mu\bar{\mu}/e\bar{e}, i}^{\text{MC}}} - \dfrac{R_{\mu\bar{\mu}/e\bar{e}, i}^{\text{SM+NP}}}{R_{\mu\mu/ee, i}^{\text{SM}}} \right)^2}{\sigma_i^2},
		\end{align}
	where $\sigma_i$ are the experimental uncertainties reported in Ref.~\cite{Sirunyan:2021khd}. Here, CMS reported an excess in electrons of about $4\sigma$, and taking into account the $q^2$ distribution, one can improve the description of the data by more than $3\sigma$ with respect to the SM~\cite{Crivellin:2021rbf}.

	ATLAS, on the other hand, did not provide the muon vs.\ electron ratio, but rather quoted limits on NP from non-resonant di-electrons in the signal region $\left [2.2, 6.0 \right]$ TeV ($\left [2.7, 6.0 \right]$ TeV) for the cases where the NP contribution is interfering constructively (destructively) with the SM. Even though their limits agree with the SM expectation within $2\sigma$, slightly more electrons than expected are observed. Here, we have to recast the bounds on the Wilson coefficients (similarly to the method described for the CMS analysis), since ATLAS studied only operators that have equal couplings to up and down quarks, which is generally not the case for the simplified models we consider.
	
	Concerning LQs, which contribute in the $t$-channel, we included the propagator according to the prescription in Ref.~\cite{Bessaa:2014jya}. For $Z^\prime$ bosons we did not include resonant LHC searches. This means implicitly that we assumed that $M$ is above the production threshold, in such a way that the bound on $M$ for $g=1$, to be derived in the next section, should be understood as a limit on $g/M$.

	\begin{table}
		\centering
		\begin{tabular}{c  c  c  c  c  c}
		\toprule
			& ATLAS destructive & ATLAS constructive & CMS & CAA & APV+$Q_\text{weak}$  \\
			& (95\%)& (95\%) & (68\%) & (68\%) & (95\%)  \\
			\midrule
			$\Phi_1$ ($\lambda_1^L$) & 2.5  & * & $-$ & $-$ & 2.6 \\
			$\Phi_1$ ($\lambda_1^R$) & 2.8  & * & $-$ & * & 4.1 \\
			$\Phi_{\tilde{1}}$ ($\lambda_1^R$) & * & 2.9  & $3.4^{+0.7}_{-0.5}$ & * & 3.9 \\
			$\Phi_{2}$ ($\lambda_2^{LR}$) & * & 4.2  & $4.3^{+0.8}_{-0.6}$ & *&3.5 \\
			$\Phi_{2}$ ($\lambda_2^{RL}$) & * & 3.8  & $4.2^{+0.9}_{-0.6}$  & * & 4.1\\
			$\Phi_{\tilde{2}}$ ($\lambda_{\tilde{2}}$) & 2.6  & * & $2.4^{+0.4}_{-0.3}$  & * & 3.9 \\
			$\Phi_{3}$ ($\lambda_{3}$) & * & 4.8  & $4.4^{+0.7}_{-0.5}$  & $4.5^{+1.0}_{-0.6}$  &4.2\\
			$V_1$ ($\kappa_{1}^{L}$) & 3.9  & * & $-$ & $-$ &5.6 \\
			$V_1$ ($\kappa_{1}^{R}$) & 4.3 & * & $3.4^{+0.4}_{-0.3}$  & * & 3.3\\
			$V_{\tilde{1}}$ ($\kappa_{\tilde{1}}$) & * & 6.6  & $7.5^{+1.7}_{-1.1}$  & * &3.7 \\
			$V_{2}$ ($\kappa_{2}^{LR}$) & 6.4  & * & $4.6^{+0.6}_{-0.5}$   & *& 8.0\\
			$V_{2}$ ($\kappa_{2}^{RL}$) & * & 4.5  & $4.5^{+0.8}_{-0.6}$  & * & 3.3\\
			$V_{\tilde{2}}$ ($\kappa_{\tilde{2}}$) & 5.4  & * & $3.7^{+0.4}_{-0.4}$  &* & 3.7 \\
			$V_{3}$ ($\kappa_{3}$) & * & 10.0  & $10.7^{+2.4}_{-1.6}$  & $6.4^{+1.4}_{-0.9}$   & 9.9\\
			\bottomrule
		\end{tabular}
		\caption{Lower limits on the LQ masses for $|\lambda|=|\kappa|=1$. For the ATLAS di-electron searches~\cite{Aad:2020otl} and PV the 95\% C.L.\ limits are given, while for CMS~\cite{Sirunyan:2021khd} and the CAA the preferred $1\sigma$ regions are shown (if the representation improves the fit with respect to the SM). Scenarios that worsen the agreement with data compared to the SM are indicated by $-$ and  $*$ marks if the corresponding observable is unaffected.}
		\label{BoundsLQ}
	\end{table}

\section{Phenomenology}
\label{sec:Pheno}

We now turn to the  phenomenological analysis, in which we will again consider the different classes of new particles separately. First we will collect  
the bounds for each representation, and then have a more detailed look at low-energy PV including a discussion of future prospects.

	\begin{figure*}
		\includegraphics[height=0.59\textwidth]{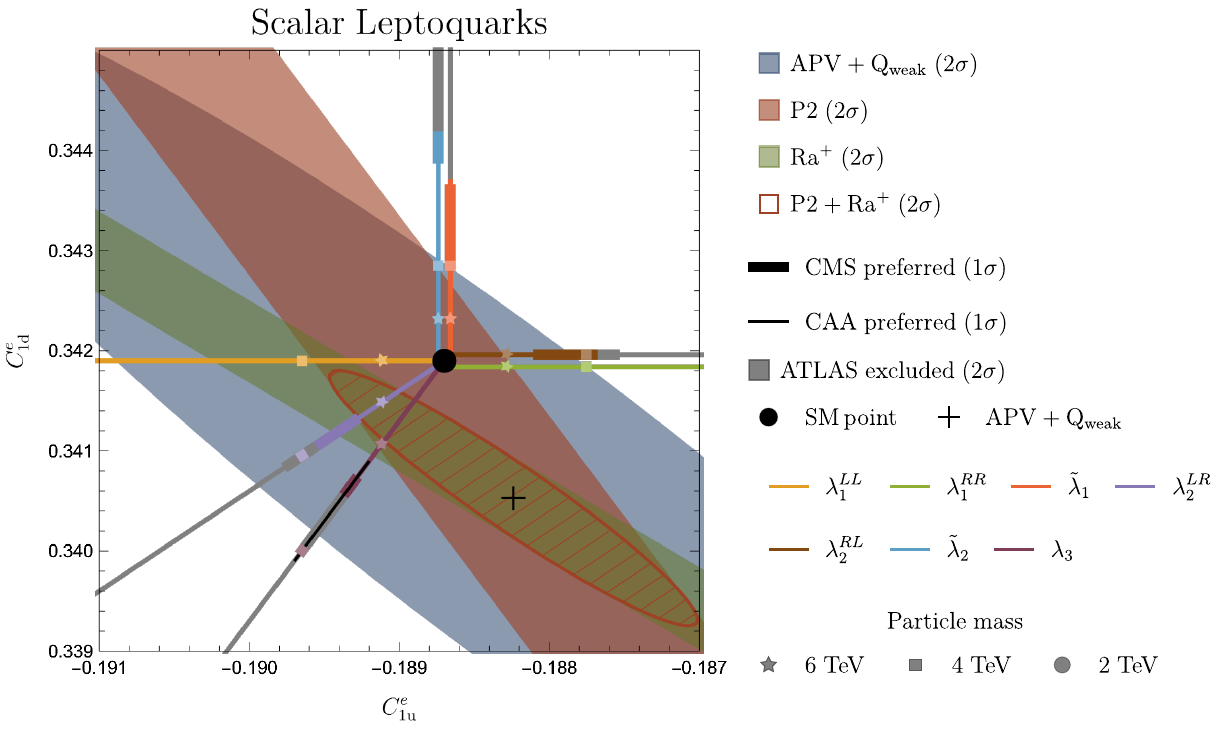} \\
		\includegraphics[height=0.59\textwidth]{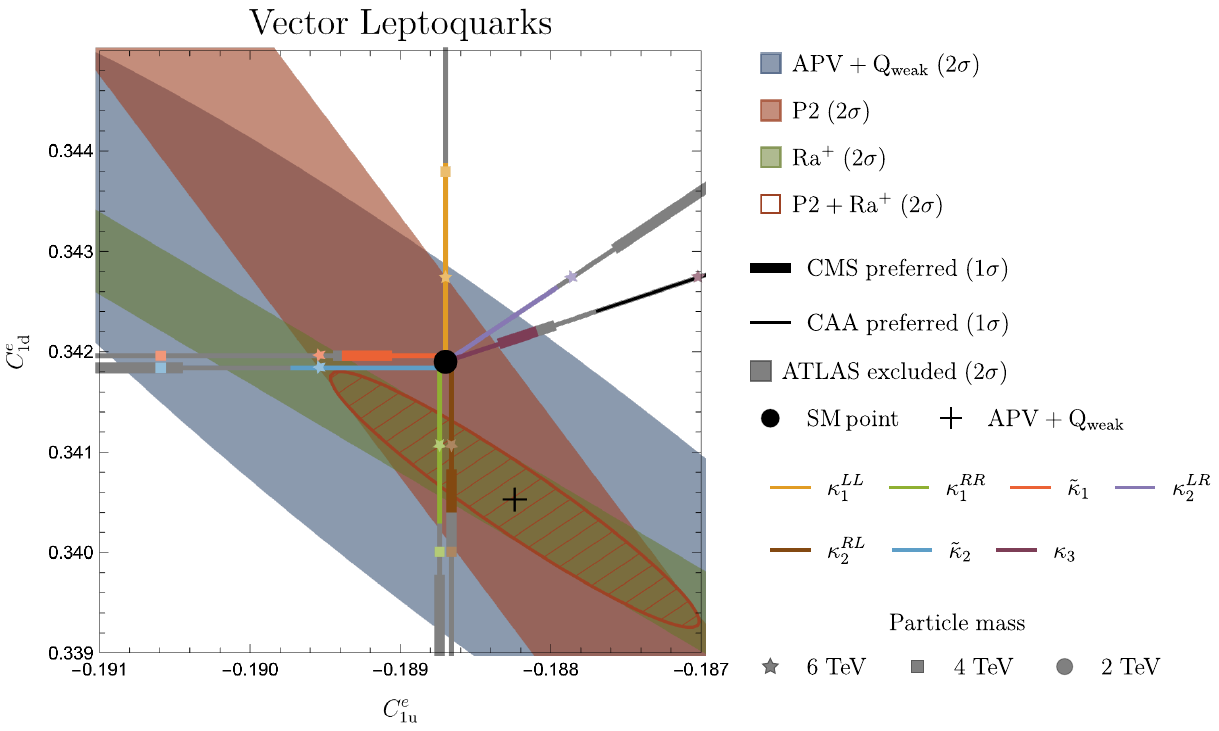}
		\caption{Parametric plot of LQ effects in the $C^e_{1u}$--$C^e_{1d}$ plane as well as the preferred regions from PV and the corresponding prospects. The gray parts of the lines are excluded by the di-electron searches of ATLAS (95\% C.L.) and the preferred regions from CMS and the CAA (both $1\sigma$) are indicated by thick and black lines, respectively. The three different values for the LQ masses ($6\,$TeV, $4\,$TeV, and $2\,$TeV), setting $\lambda$, $\kappa = 1$, are indicated by markers of different shapes, the cross denotes the best-fit point of APV and $Q_\text{weak}$, and the black circle the SM point. }
		\label{fig:LQ_e_plot}
	\end{figure*}

	\subsection{Leptoquarks}
	
	First-generation LQs were studied in Refs.~\cite{Davidson:1993qk,Bessaa:2014jya,Bansal:2018eha,Raj:2016aky,Schmaltz:2018nls,Crivellin:2021egp}. The summary of the bounds and preferred values for the LQ masses from LHC searches, PV, and the CAA is given in Table~\ref{BoundsLQ} and shown in Fig.~\ref{fig:LQ_e_plot} in the $C^e_{1u}$--$C^e_{1d}$ plane. One can see that the ATLAS bounds from non-resonant di-lepton searches are very stringent, but still allow for an explanation of the CMS excess for the LQ representations with the couplings $\tilde{\lambda}_1, \lambda_2^{LR}, \lambda_2^{RL}, \lambda_3, \tilde{\kappa}_1, \kappa_2^{RL}$, and $\kappa_3$ that give rise to constructive interference with the SM. While an explanation of the CAA with $\kappa_3$ is in tension with the ATLAS and APV+$Q_\text{weak}$ measurements, the scalar triplet with the coupling $\lambda_3$ could address the anomaly.\footnote{In principle, it would lead to too large effects in $D^0$--$\bar D^0$ mixing, but these bounds can be avoided by tuning against the (unknown) SM contribution or other NP effects.} While the current limits from PV experiments are only in some cases competitive with, or superior to, LHC limits, the experimental prospects look more promising. In fact, the Ra$^{+}$ and P2 experiments would be able to distinguish different representations and, e.g., favor $\lambda_2^{LR}$ and $\kappa_2^{RL}$ as solutions to the CMS excess, if the current central values for $C_{1u}^e$ and $C_{2u}^e$ were confirmed. 
	Regarding the PV experiments involving neutrinos, not even the 1\% accuracy projection for an Ar-COHERENT-type experiment can compete with the direct search limits, partly because only the fraction $f_e \approx 30\%$ of electron neutrinos in the overall neutrino flux can be used to search for first-generation LQs.

		\begin{table}
			\centering
		\begin{tabular}{c  c  c  c  c  c  c}
		\toprule
			& \multicolumn{2}{c}{ATLAS (95\%)} & CMS & CAA & \multicolumn{2}{c}{APV+$Q_\text{weak}$ (95\%)}\\
			& destr. & constr.  & (68\%) & (68\%) &   \\
			\midrule
			$X^I$($g^\ell_Xg^q_X$) & ($+$) 2.1  &($-$) 4.3  &($-$) $5.3^{+1.1}_{-1.1}$  & ($-$) $4.5^{+1.0}_{-0.6}$ &($+$) 1.4 & ($-$) 1.4 \\
			$Z^{\prime}$ ($g^\ell g^q$) & ($-$) 4.5  &($+$) 7.3  &($+$) $7.1^{+1.5}_{-0.9}$ &*&($-$) 5.0 & ($+$) 8.0\\
			$Z^{\prime}$ ($g^\ell g^u$) & ($-$) 4.2  &($+$) 6.5  &($+$) $6.3^{+1.3}_{-0.8}$  &*&($-$) 3.8 & ($+$) 5.8\\
			$Z^{\prime}$ ($g^\ell g^d$) & ($+$) 3.6 &($-$) 5.1   &($-$) $4.8^{+0.8}_{-0.6}$  &*&($+$) 3.4 & ($-$) 5.6\\
			$Z^{\prime}$ ($g^\ell g^u=g^\ell g^d$) & ($-$) 4.7  &($+$) 6.9  &($+$) $6.4^{+1.2}_{-0.8}$  &*&($-$) 5.0 & ($+$) 8.0\\
			$Z^{\prime}$($g^\ell g^u=-g^\ell g^d$)&($-$) 4.6  &($+$) 7.2  & ($+$) $7.1^{+1.4}_{-0.9}$ &*&($-$) 2.8 &($+$) 2.8\\
			$Z^{\prime}$ ($g^eg^q$) & ($-$) 4.7  &($+$) 6.9  &($+$) $6.5^{+1.2}_{-0.8}$ &*&($-$) 4.9 & ($+$) 8.0\\
			$Z^{\prime}$ ($g^eg^u$) & ($-$) 4.2  &($+$) 7.1  &($+$) $7.6^{+1.8}_{-1.1}$  &*&($-$) 3.7 & ($+$) 5.8\\
			$Z^{\prime}$ ($g^eg^d$) & ($+$) 3.6 &($-$) 5.3   &($-$) $5.2^{+1.0}_{-0.6}$  &*&($+$) 3.3 & ($-$) 5.6\\
			$Z^{\prime}$ ($g^eg^u=g^eg^d$) & ($-$) 4.7  &($+$) 7.3  & ($+$) $7.2^{+1.5}_{-0.9}$  &*&($-$) 4.9 & ($+$) 8.0\\
			$Z^{\prime}$ ($g^eg^u=-g^eg^d$) & ($-$) 4.6  &($+$) 5.4  & ($+$) $8.7^{+1.9}_{-1.3}$ &* &($-$) 2.8 & ($+$) 2.8\\
			\bottomrule
		\end{tabular}
		\caption{Lower limits on the masses of VBs for $|g^e|=|g^q|=1$ extracted from the ATLAS di-electron analysis and PV, as well as the mass ranges preferred by the CAA and CMS data ($1\sigma$). The $*$ mark indicates if the corresponding observable is unaffected. The signs in the brackets denote the sign of the product $g^eg^q$. The notation for the couplings is as in Eq.~\eqref{Lagr_VB}, with only the couplings as indicated in brackets in the first column non-vanishing.}
		\label{BoundsVB}
	\end{table}

		\begin{figure*}
	\centering
		\includegraphics[width=0.99\textwidth]{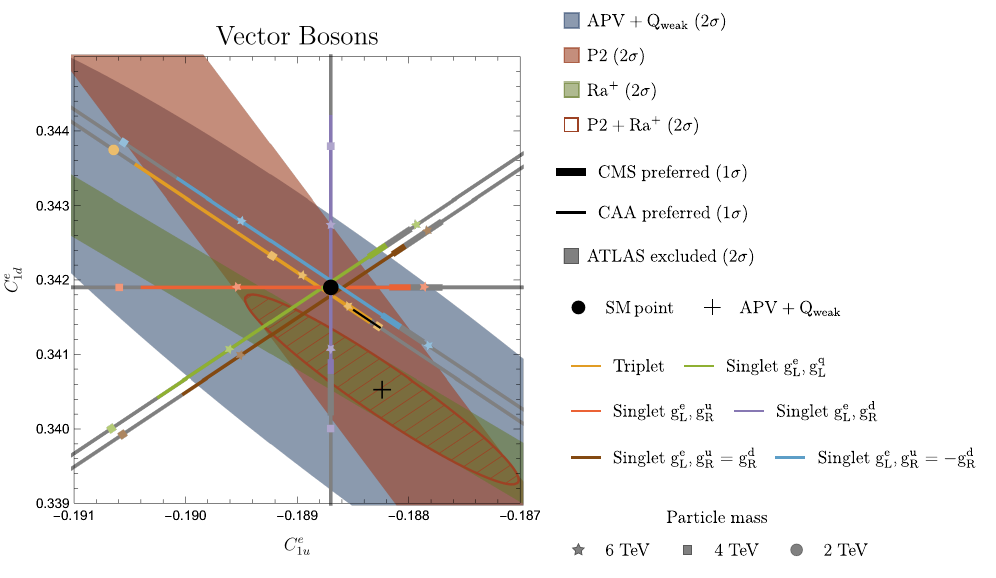} \vspace{20px}\\
		\includegraphics[width=0.99\textwidth]{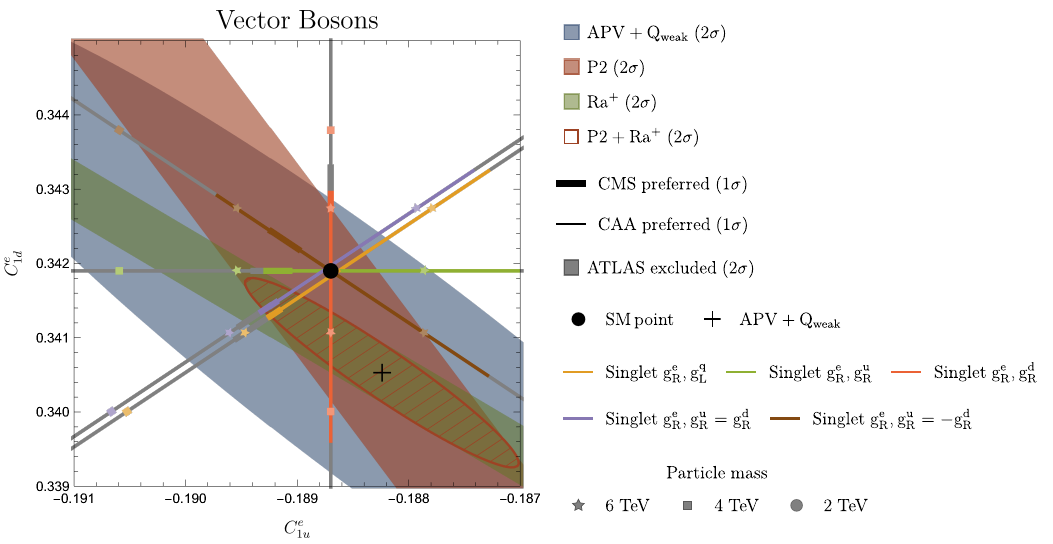}
		\caption{Parametric plot of VB effects in the $C^e_{1u}$--$C^e_{1d}$ plane, as well as the preferred regions from PV and the corresponding prospects. The gray parts of the lines are excluded by the di-electron searches of ATLAS (95\% C.L.) and the preferred regions from CMS and the CAA (both $1\sigma$) are indicated by thick and black lines, respectively. The three different values for the VB masses ($6\,$TeV, $4\,$TeV, and $2\,$TeV), setting $\lambda$, $\kappa = 1$, are indicated by markers of different shapes, the cross denotes the best-fit point of APV and $Q_\text{weak}$, and the black circle the SM point. } \vspace{30px}
		\label{fig:VB_plot}
\end{figure*}
		
	\subsection{Vector Bosons}
	
	For VB we consider two cases, 
	an $SU(2)_L$ singlet ($Z^\prime$) and an $SU(2)_L$ triplet ($X^I$). The $Z^\prime$ couplings to left- and right-handed quarks and leptons are all independent,\footnote{For a detailed analysis of the $SU(2)_L$ singlet coupling only to leptons see Ref.~\cite{Buras:2021btx}. EW constraints on VBs have been studied in Refs.~\cite{deBlas:2012qp,delAguila:2010mx} and PV in Refs.~\cite{Casalbuoni:1999mw,Long:2018fud,DAmbrosio:2019tph,Dev:2021otb}.} and therefore it is evident from Eq.~\eqref{eq:VBsingletcouplings} that various combinations of Wilson coefficients can be generated. In our phenomenological analysis we study the benchmark scenarios given in Table~\ref{BoundsVB}, which are shown in the $C_{1u}^e$--$C_{1d}^e$ plane in Fig.~\ref{fig:VB_plot}. Even though the ATLAS bounds are more stringent in case of constructive interference with the SM, allowing only masses above ($5$--$7$)TeV (if the relevant BSM couplings are fixed to unity), we find that each scenario is able to explain the CMS excess at the $1\sigma$ level. While the $Z^\prime$ does not affect the CAA, $X^I$ generates a contribution to $C_{\ell q}^{(3)}$, which can, for constructive interference, explain the CAA and the CMS excess simultaneously (see Fig.~\ref{fig:VB_plot}). Concerning PV experiments, the current APV and $Q_\text{weak}$ measurements yield already competitive bounds for some cases, which would improve significantly with P2 and APV with Ra$^+$. Assuming the central value to be the same as from current APV and $Q_\text{weak}$ experiments, one could, e.g., disfavor $X^I$ and only two scenarios for the $Z^\prime$ would be preferred. 
	
	\begin{table}
		\centering
	\begin{tabular}{ c  c  c  c}
	\toprule
		& EWPO & CAA & APV+$Q_\text{weak}$  \\
		& (95\% / 68\%) & (68\%) & (95\%)  \\
		\midrule
		$U$  & $3.6$ & $4.4^{+1.0}_{-0.6}$  & $3.8$  \\
		$D$  & $4.1$ & $4.4^{+1.0}_{-0.6}$  & $2.4$  \\
		$Q_1 (\xi^{u_1})$  & $2.4$ & * & $2.7$  \\
		$Q_1 (\xi^{d_1})$  & $1.7$  & *& $3.8$ \\
		$Q_5$  & $1.6^{+3.3}_{-1.2}$ & *& $2.4$  \\
		$Q_7$  & $2.4^{+5.0}_{-1.7}$  & *& $3.9$ \\
		$T_1$  & $1.5^{+3.6}_{-1.1}$ & $-$ & $2.2$  \\
		$T_2$ & $1.9^{+4.0}_{-1.5}$ & $-$ & $3.2$  \\
		\bottomrule
	\end{tabular}
	\caption{Ranges or lower limits for the masses of VLQ for $|\xi|=1$ extracted from EWPO and PV (95\% C.L.). The first two representations can explain the CAA at the $1\sigma$ level for the given mass rage, while $-$ indicates that the fit is worsened with respect to the SM, and * denotes that the CAA is not affected. }
	\label{BoundsVLQ}
\end{table}

	\begin{figure*}
		\begin{center}
			\raisebox{0.6em}{\includegraphics[height=0.59\textwidth]{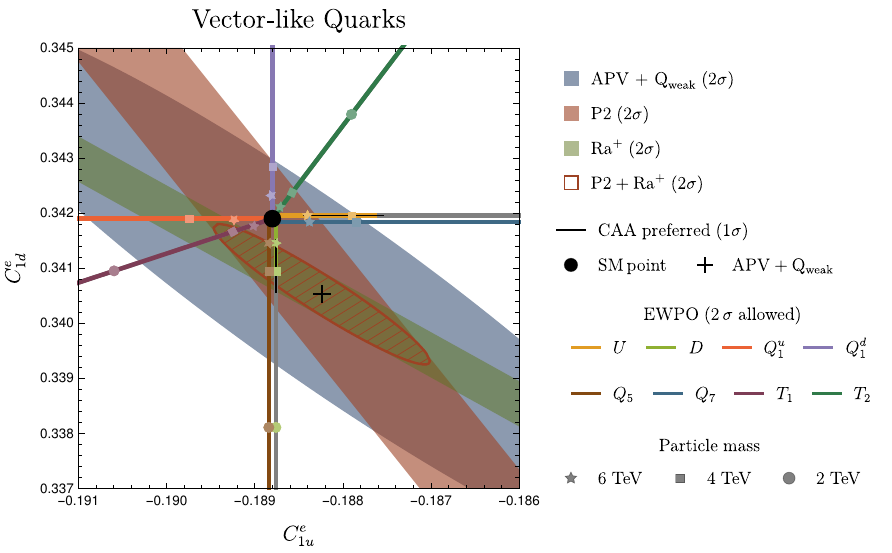}}
			\raisebox{0.6em}{\includegraphics[height=0.59\textwidth]{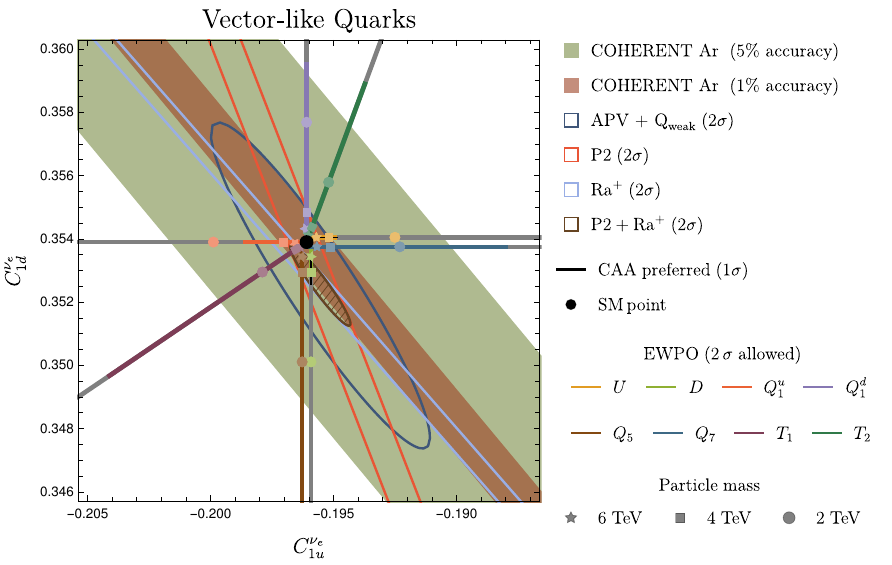}}
			\caption{Parametric plot of VLQ effects in the $C^e_{1u}$--$C^e_{1d}$ and $C^{\nu_e}_{1u}$--$C^{\nu_e}_{1d}$ planes, as well as the preferred regions from PV and the corresponding prospects. The gray parts of the lines are excluded by the di-electron searches of ATLAS (95\% C.L.) and the preferred regions from CMS and the CAA (both $1\sigma$) are indicated by thick and black lines, respectively. The three different values for the VB masses ($6\,$TeV, $4\,$TeV, and $2\,$TeV), setting $\lambda$, $\kappa = 1$, are indicated by markers of different shapes.
			}
			\label{fig:VLQ_plot}
		\end{center}
	\end{figure*}	
	
		\begin{figure*}
		\begin{center}
			\includegraphics[height=0.6\textwidth]{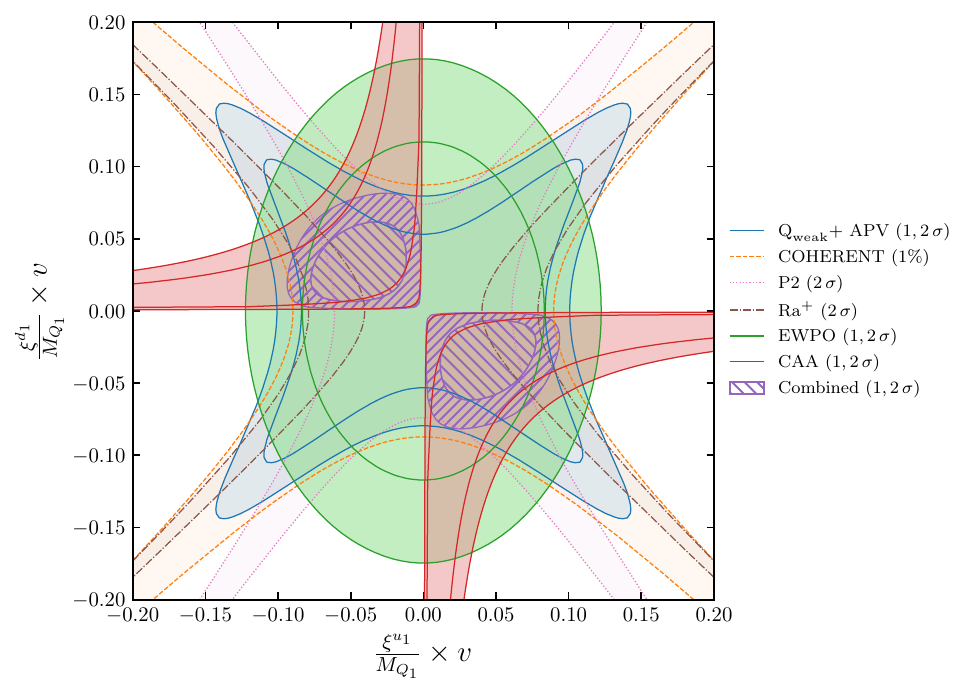}
			\caption{Preferred regions in the $v\xi^{u_1}/M_{Q_1} $--$ v\xi^{d_1}/M_{Q_1}$ plane for the $Q_1$ VLQ. ``Combined'' stands for the statistical combination of $Q_\text{weak}$, APV, EWPO, and the CAA.}
			\label{fig:VLQ_Q1_plot}
		\end{center}
	\end{figure*}
	
	\subsection{Vector-like Quarks}
	
We first note that only the three representations $U$, $D$, or $Q_1$ (with couplings to both up and down quarks) can explain the CAA~\cite{Belfatto:2019swo,Cheung:2020vqm,Belfatto:2021jhf,Branco:2021vhs} due to their mixing with the SM quarks, while, on the other hand, $T_1$ or $T_2$ worsen the CAA.
EWPO places multi-TeV mass limits on the VLQs $U$, $D$, and $Q_1$ (for couplings fixed to unity), which come close to ruling out the best-fit regions from the CAA for $U$ and $D$, while there is actually a small ($1\sigma$) preference for the other representations ($Q_{5,7}$ and $T_{1,2}$) driven by $R^0_{e}$ and $R^0_{\mu}$. All these bounds are collected in Table~\ref{BoundsVLQ} and shown in Fig.~\ref{fig:VLQ_plot}. 
While there have been several direct searches for VLQs coupling to first-generation quarks, the mass limits are $\sim 1 \TeV$ \cite{ATLAS:2011tvb,ATLAS:2012apa,ATLAS:2015lpr,CMS:2017asf} (or perhaps as high as $\sim 1.4 \TeV$ in certain regions of parameter space for $D$ or $T_2$) and so are currently weaker than any of the other observables considered in this paper.

Looking back at Table~\ref{BoundsVLQ} and Fig.~\ref{fig:VLQ_plot} we also see that $Q_\text{weak}$ and APV currently provide in several cases even better bounds than EWPO. Since the $Q_1$ VLQ has two free parameters, as it can couple to both up and down quarks, we show the full parameter space in Fig.~\ref{fig:VLQ_Q1_plot}.
Concerning future prospects, a 1\% accuracy is needed for an Ar-COHERENT-type experiment to be relevant, but P2 and Ra$^+$ would (assuming that the current central value is confirmed) still allow the CAA to be explained by either the $D$ VLQ or a $Q_1$ VLQ coupling to both $u$ and $d$ quarks.

\subsection{Vector-like Leptons}

VLLs coupling to first-generation leptons are strongly constrained by EWPO from LEP and the LHC~\cite{delAguila:2008pw,Antusch:2014woa,deGouvea:2015euy,Fernandez-Martinez:2016lgt,Chrzaszcz:2019inj,Das:2020uer,Crivellin:2020ebi,Manzari:2021prf} as well as direct searches~\cite{L3:2001xsz,ATLAS:2019kpx,CMS:2019hsm}. The direct search bounds are $\sim700\GeV$, $\sim450\GeV$, and $\sim150\GeV$ for doublets, triplets, and singlets respectively, while EWPO bounds are even stronger: doublets and triplets to masses above $\sim 5\TeV$ and $\sim 4\TeV$ respectively, while there is some preference for the $SU(2)_L$ singlet VLL at around $7\TeV$ (for a coupling fixed to unity), driven by the discrepancy in $R^0_e$, $R^0_\mu$, and $A_e$. They also cannot substantially improve the CAA, since only $V_{us}$ as determined from $K_{\ell 3}$ decays is affected, while all the other main determinations of $V_{us}$ and $V_{ud}$ are unchanged with only first-generation lepton couplings (see Refs.~\cite{Kirk:2020wdk,Coutinho:2019aiy,Manzari:2020eum,Crivellin:2020lzu,Crivellin:2020ebi,Manzari:2021prf} for VLLs explaining the CAA with coupling to different lepton generations.). Furthermore, the EWPO bounds are so strong that PV cannot compete here, even taking into account possible future improvements.

	\begin{figure*}
		\centering
		\includegraphics[width=0.95\textwidth]{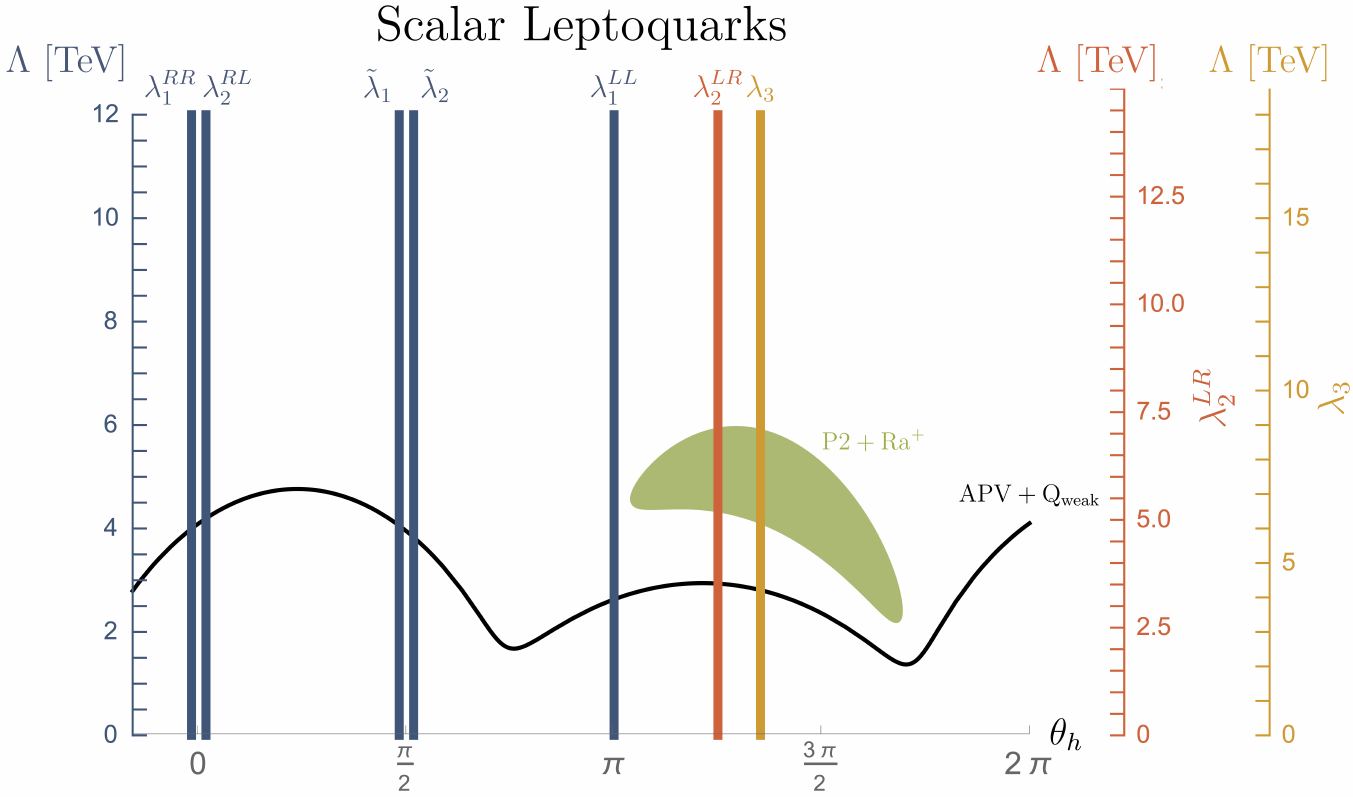} \vspace{20px} \\
		\includegraphics[width=0.95\textwidth]{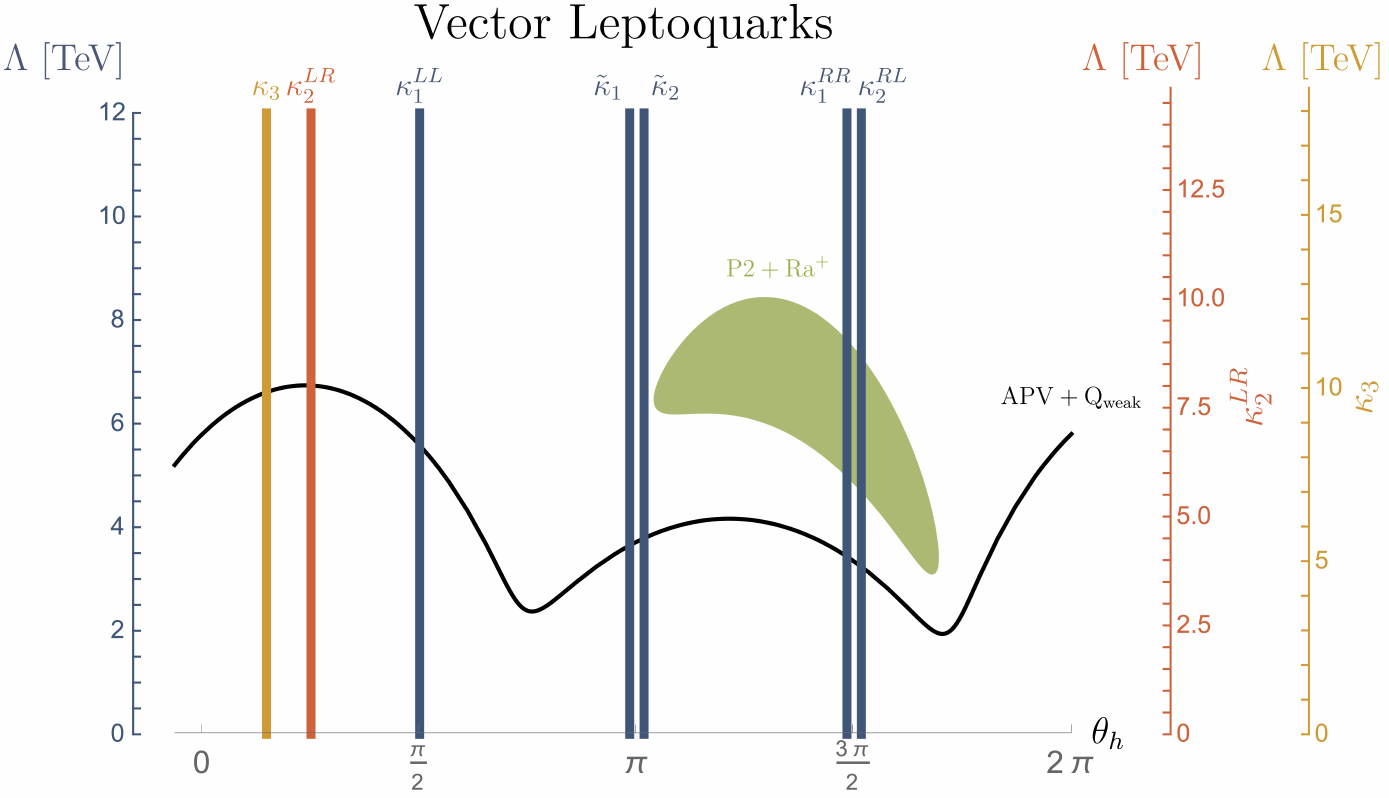}
		\caption{95\% C.L.\ exclusion limits from $\text{APV}+Q_{\text{weak}}$ for the different LQ representations. The angle $\theta_h$ refers to the phase of the NP contribution in the $C_{1u}^e$--$C_{1d}^e$ plane, i.e., $\theta_h = \arctan \big( C_{1d}^{e, \text{NP}}/C_{1u}^{e, \text{NP}}\big)$. The couplings $\lambda_2^{LR}, \kappa_2^{LR}$ and $\lambda_3, \kappa_3$ require a different normalization, therefore the different scales on the right apply when reading off the scale $\Lambda$ (corresponding to the LQ mass). The black line refers to the current limit from $\text{APV}+Q_{\text{weak}}$, the green region to the preferred space assuming $\text{P2}+\text{Ra}^+$ precision and no change in the central values.}
		\label{fig:LQconclusions}
	\end{figure*}

	\subsection{Limits and Prospects from Low-Energy Parity Violation}
	\label{sec:LQ_PV}
	
	NP limits from low-energy PV have already been formulated in terms of the NP scale $\Lambda$ for benchmark models in Refs.~\cite{Androic:2018kni,Wang:2014guo}. Here, we compare our results to the presentation in Ref.~\cite{Androic:2018kni}, where the combined limits from $Q_\text{weak}$ and APV are given as a function of the angle $\theta_h = \arctan \big( C_{1d}^{e, \text{NP}}/C_{1u}^{e, \text{NP}}\big)$, as shown in Fig.~\ref{fig:LQconclusions}. The first conclusion from the simplified-model analysis is that only discrete values in $\theta_h$ arise, as indicated by the vertical lines in Fig.~\ref{fig:LQconclusions}. Depending on the angle, the limits on the respective LQ masses can vary substantially, between $2.5\TeV$ and $10\TeV$, see also Table~\ref{BoundsLQ}, as reflects the fact that the different simplified models define a different trajectory in the $C_{1u}^e$--$C_{1d}^e$ plane and thus a different intersection with the current exclusion ellipse, see Fig.~\ref{fig:LQ_e_plot}.  The projection for $\text{P2}+\text{Ra}^+$ then emphasizes the discriminating power of the combined analysis: if the central values remained at the current $\text{APV}+Q_{\text{weak}}$ best-fit point, but errors were reduced as projected, the combined data would prefer the green region in Fig.~\ref{fig:LQconclusions}, and thus eliminate most of the simplified models as suitable candidates.  
	
	The analysis in our paper is focused on the $C_{1q}$ coefficients, but we do keep the full dependence also on $C_{2q}$ (in the case of $Q_\text{weak}$) and studied the impact of the PVDIS constraints including projected limits from SoLID, see Sec.~\ref{sec:PVDIS}. Scenarios in which $C_{2q}$ dominate are possible in $Z'$ models upon tuning the couplings accordingly~\cite{Buckley:2012tc,Gonzalez-Alonso:2012aib}, but the same coefficients can also be probed in the LHC di-lepton searches. As show in Ref.~\cite{Boughezal:2021kla} in the framework of SMEFT, the estimated SoLID sensitivity to NP is at most of the same order as current LHC searches, and the simplified-model analysis yields the same conclusion. However, in the combination with P2 and di-lepton searches, it could be helpful in distinguishing different NP scenarios.

	\begin{figure*}[t]
	\centering
	\includegraphics[width=0.99\textwidth]{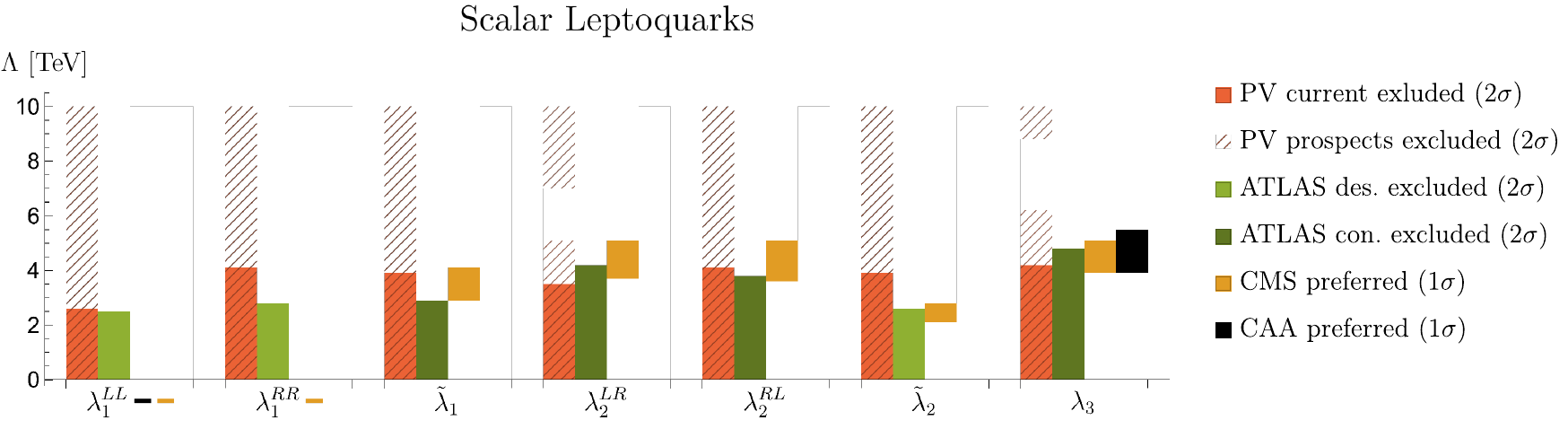} \vspace{10px} \\
	\includegraphics[width=0.99\textwidth]{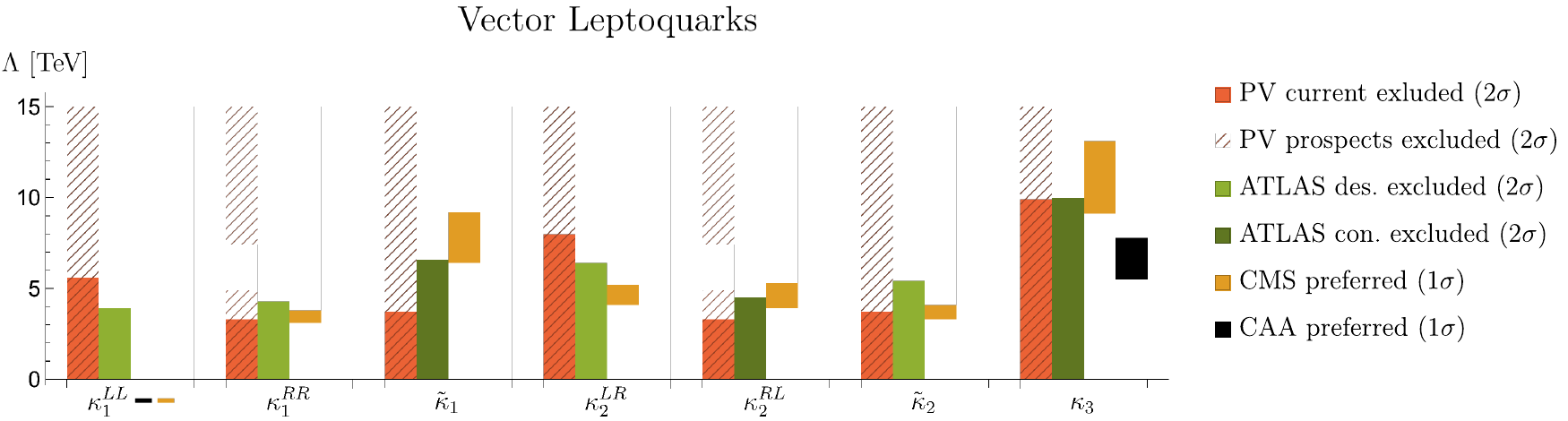}\vspace{10px} \\
	\includegraphics[width=0.99\textwidth]{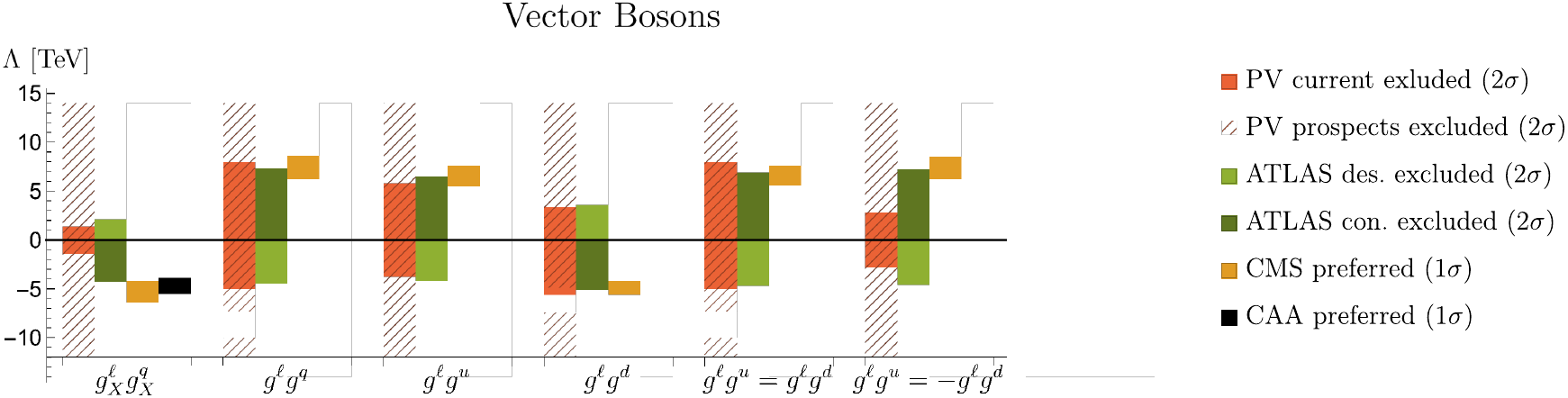} \vspace{10px} \\
	\includegraphics[width=0.99\textwidth]{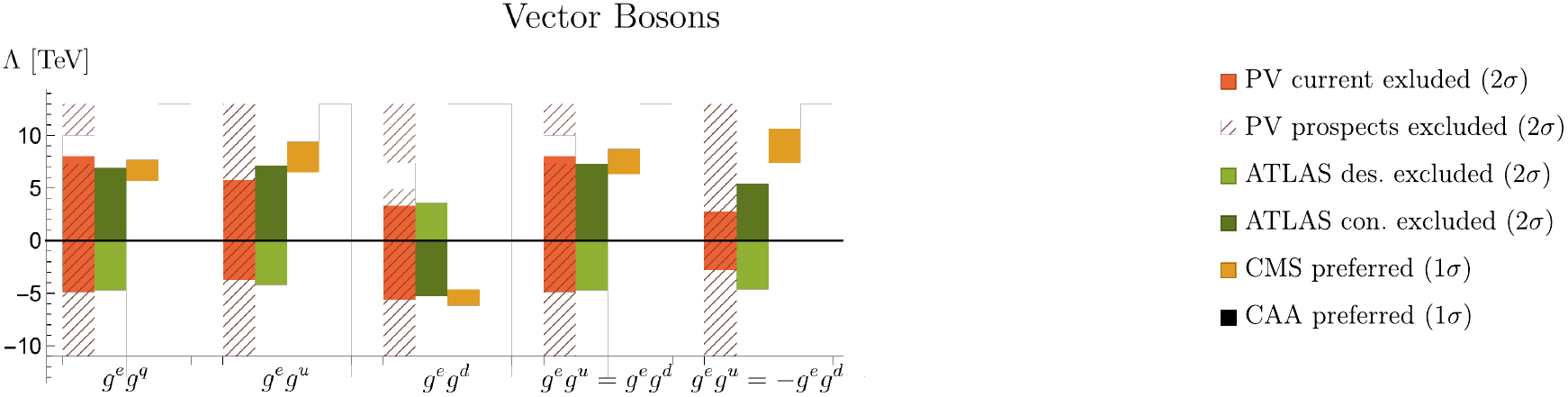}
	\caption{Graphical representation of the limits on the LQ and VB mass scale $\Lambda$ for $\left | \lambda \right | = \left | \kappa \right | = 1$ and $\left | g^e \right | = \left | g^q \right | = 1$, respectively. Scenarios that worsen the agreement with CAA (CMS) data compared to the SM are indicated with black (yellow) minus signs next to the coupling label. The hatched regions indicate the P2 and Ra$^+$ exclusion prospects, assuming that the current central value remains unchanged. }
	\label{fig:conclusions1}
\end{figure*}

	\begin{figure*}
		\raisebox{0.6em}{\includegraphics[height=0.29\textwidth]{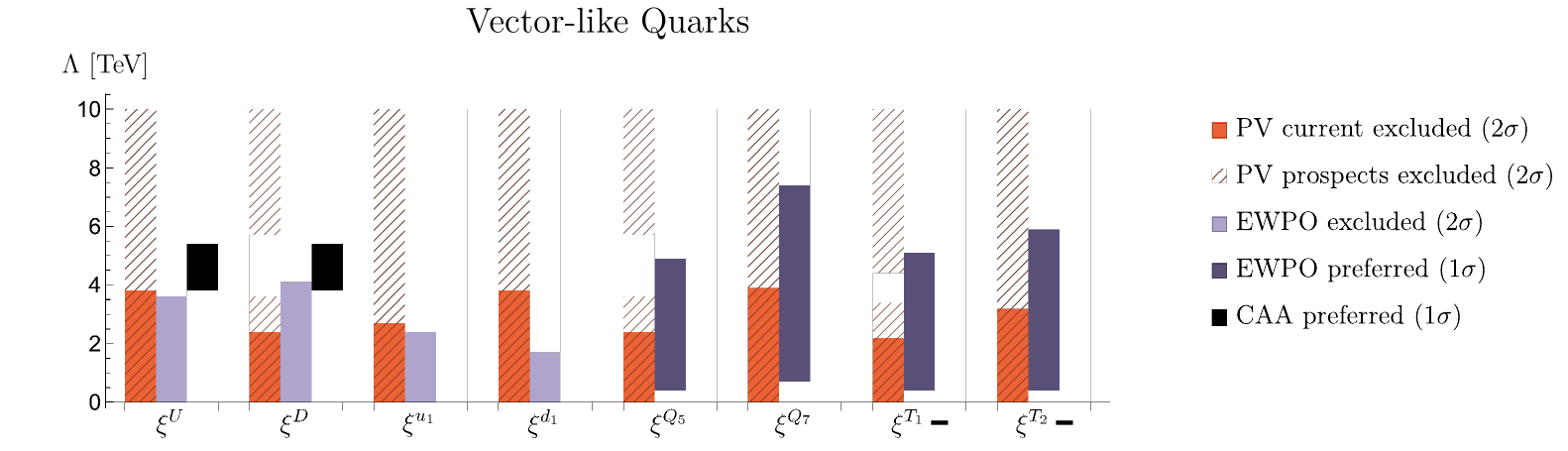}}\\
		\raisebox{0.6em}{\includegraphics[height=0.29\textwidth]{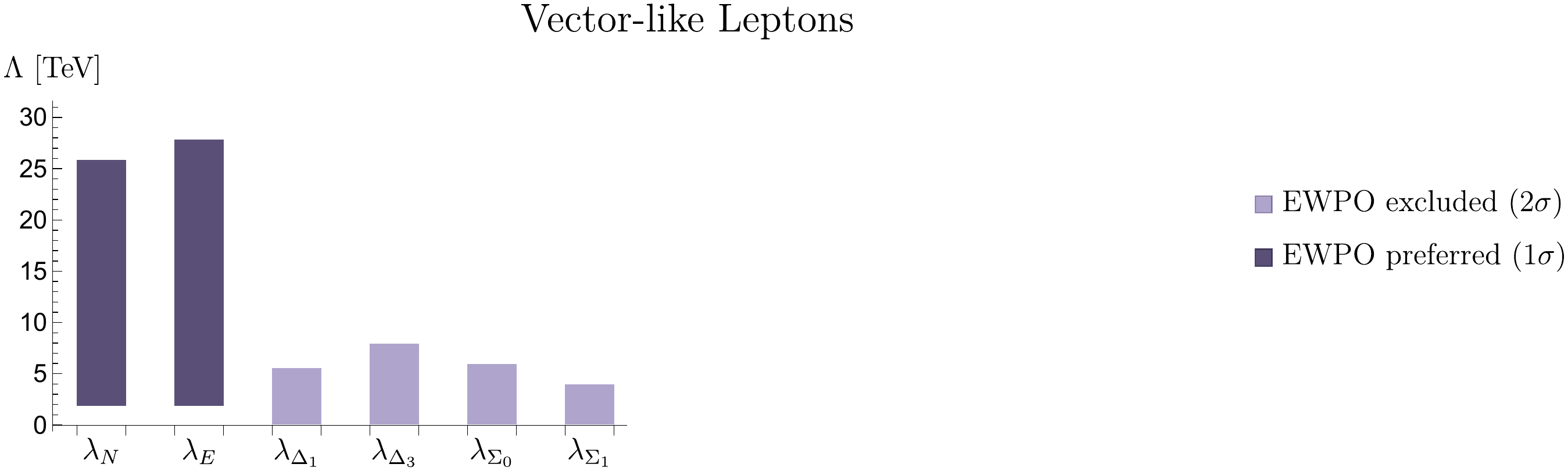}}
		\caption{Graphical representation of the limits on the VLQ (VLL) mass scale $\Lambda$ for $|\xi|=1$ ($|\lambda|=1$). The hatched regions indicate the P2 and Ra$^+$ exclusion prospects, assuming that the current central value remains unchanged. For VLL only the bounds from EWPO are relevant. }
		\label{fig:conclusions2}
\end{figure*}

	\section{Summary and Conclusions}
	\label{sec:Conclusions}

	NP related to first-generation fermions is not only very constrained, but also particularly interesting due to the large variety of related observables, allowing one to probe many complementary aspects of possible SM extensions. More recently, the observed deficit in the first-row and 
	-column CKM unitarity, known as the Cabibbo angle anomaly (CAA), as well as the excess in non-resonant di-electron searches observed by CMS,
	further motivate the study of NP effects in the first generation. Since, due to the large number of couplings in SMEFT, fully model-independent relations among different processes are still scarce, the identification of such correlations suggests the consideration of simplified models, especially the four classes that can give rise to modified 2-quark--2-lepton interactions at tree level (after EW symmetry breaking): leptoquarks (LQs), QCD-neutral vector bosons (VBs), vector-like quarks (VLQs), and vector-like leptons (VLLs). 
	
	After performing the matching of these models onto the dimension-6 $SU(2)_L$ gauge invariant SMEFT, we analyzed all relevant observables related to these operators, including $\beta$ decays, EWPO, LHC bounds, and low-energy PV.
	In particular, we provided master formulae for PVES, Eq.~\eqref{AVAA}, and CE$\nu$NS, Eq.~\eqref{CEvNS_SM}, that express the observables directly in terms of the short-distance Wilson coefficients and the respective hadronic matrix elements. 
	Our main results are summarized in Fig.~\ref{fig:conclusions1} and Fig.~\ref{fig:conclusions2}, illustrating the complementarity of the different classes of observables. For instance,  the LHC bounds obtained from the CMS and ATLAS  non-resonant di-electron searches only apply to LQs and VBs, while these extensions are in general poorly constrained by EWPO. VLLs are by far best bounded by EWPO, which is also the case for several VLQ representations. LQs, VBs, and VLQs can have a significant impact on PV, but the actual  constraint depends on the representation, and, similarly,  only a subset of the representations affects $\beta$ decays and thus the CAA.

	Concerning PV, different representations predict different trajectories in the $C_{1q}^e$ and $C_{1q}^\nu$ planes departing from the SM point, which leads to different bounds given that the present exclusion limits are not symmetric. This is illustrated for LQs in 
	Fig.~\ref{fig:LQconclusions}, as a function of the phase $\theta_h = \arctan \big( C_{1d}^{e, \text{NP}}/C_{1u}^{e, \text{NP}}\big)$. Furthermore, future experimental improvements
	could allow one to disentangle different representations, most notably with next-generation PV experiments P2 and Ra$^+$, as shown in Figs.~\ref{fig:LQconclusions}, ~\ref{fig:conclusions1}, and~\ref{fig:conclusions2}. In addition, we find that for VLQs larger effects in PV than in the other simplified NP models are possible, in such a way that future prospects for CE$\nu$NS also promise discriminating power. 
	
	Altogether, Figs.~\ref{fig:conclusions1} and~\ref{fig:conclusions2} show 
	the complementarity of low-energy precision observables and LHC searches in constraining NP related to first-generation quarks and leptons.
	If, further, the CAA and/or the CMS excess in di-electrons were confirmed, PV and EWPO could be used to disentangle different NP models even in case the masses were so heavy that a direct discovery would require a center-of-mass energy only available at future colliders.

\acknowledgments 
	
	We thank P.~King and J.~Erler for helpful correspondence on Refs.~\cite{Androic:2018kni,Erler:2013xha}, and Greg Smith for comments on the manuscript. Financial support by the Swiss National Science Foundation, Project Nos.\ PP00P2\_176884 (A.C.\ and C.A.M.) and PCEFP2\_181117 (M.H.), and the ``Excellence Scholarship \& Opportunity Programme'' of the ETH Z\"urich (L.S.) is gratefully acknowledged. A.C.\ thanks CERN for the support via the Scientific Associate program.
	M.K.\ was financially supported by MIUR (Italy) under a contract PRIN 2015P5SBHT and by INFN Sezione di Roma La Sapienza and partially supported by the ERC-2010 DaMESyFla Grant Agreement Number: 267985.

	\bibliographystyle{apsrev4-1_mod}
	\bibliography{BIB}
	
\end{document}